\documentclass[12pt]{article}

\usepackage{amssymb}
\usepackage{graphicx}
\usepackage{mathptmx}
\usepackage{microtype}
\usepackage{listings}
\usepackage[pdftex,dvipsnames]{xcolor}
\usepackage{xspace}
\usepackage{xargs}
\usepackage{xcolor,colortbl}
\usepackage{mfirstuc}
\usepackage{multirow}
\usepackage{url}
\usepackage[colorinlistoftodos,shadow]{todonotes}
\usepackage{pdflscape}
\usepackage{colortbl}
\usepackage{hhline}
\usepackage{amsmath}
\usepackage[shortcuts]{extdash}
\usepackage[export]{adjustbox} 
\usepackage{changepage}
\usepackage{subfig}
\usepackage[colorlinks = true,
            linkcolor = darkblue,
            urlcolor  = darkred,
            citecolor = darkblue,
            anchorcolor = blue]{hyperref}

\usepackage{lineno}

\graphicspath{{graphics/}}

\usepackage{array}

\definecolor{darkred}{rgb}{0.5,0.0,0.0}
\definecolor{darkgreen}{rgb}{0.0,0.4,0.0}
\definecolor{darkblue}{rgb}{0.0,0.0,0.5}

\newcommand{\mytodo}[1]%
{%
\addcontentsline{toc}{subsection}{\protect \textbf{TODO: }#1}%
\textbf{TODO: #1}%
}

\newtheorem{mydef}{Definition}

\definecolor{bananamania}{rgb}{0.98, 0.91, 0.71}

\newcolumntype{P}[1]{>{\centering\arraybackslash}p{#1}}
\newcolumntype{M}[1]{>{\centering\arraybackslash}m{#1}}

\newcolumntype{L}[1]{>{\raggedright\let\newline\\\arraybackslash\hspace{0pt}}p{#1}}
\newcolumntype{C}[1]{>{\centering\let\newline\\\arraybackslash\hspace{0pt}}p{#1}}
\newcolumntype{R}[1]{>{\raggedleft\let\newline\\\arraybackslash\hspace{0pt}}p{#1}}

\newcolumntype{+}{!{\vrule width 2pt}}

\newlength\savedwidth

\newcommand\thickhline{\noalign{\global\savedwidth\arrayrulewidth\global\arrayrulewidth 2pt}%
\hline
\noalign{\global\arrayrulewidth\savedwidth}}

\newcommand{\orcid}[1]{\href{https://orcid.org/#1}{\textcolor[HTML]{A6CE39}{\aiOrcid}}}

\usepackage{scalerel}
\usepackage{tikz}
\usetikzlibrary{svg.path}

\definecolor{orcidlogocol}{HTML}{A6CE39}
\tikzset{
  orcidlogo/.pic={
    \fill[orcidlogocol] svg{M256,128c0,70.7-57.3,128-128,128C57.3,256,0,198.7,0,128C0,57.3,57.3,0,128,0C198.7,0,256,57.3,256,128z};
    \fill[white] svg{M86.3,186.2H70.9V79.1h15.4v48.4V186.2z}
                 svg{M108.9,79.1h41.6c39.6,0,57,28.3,57,53.6c0,27.5-21.5,53.6-56.8,53.6h-41.8V79.1z M124.3,172.4h24.5c34.9,0,42.9-26.5,42.9-39.7c0-21.5-13.7-39.7-43.7-39.7h-23.7V172.4z}
                 svg{M88.7,56.8c0,5.5-4.5,10.1-10.1,10.1c-5.6,0-10.1-4.6-10.1-10.1c0-5.6,4.5-10.1,10.1-10.1C84.2,46.7,88.7,51.3,88.7,56.8z};
  }
}

\newcommand\orcidicon[1]{\href{https://orcid.org/#1}{\mbox{\scalerel*{
\begin{tikzpicture}[yscale=-1,transform shape]
\pic{orcidlogo};
\end{tikzpicture}
}{|}}}}

\begin{document}

\title{
\capitalisewords{Dataset search in biodiversity research:} \\
\capitalisewords{Do metadata in data repositories reflect} \\
\capitalisewords{scholarly information needs?}}

\author{\normalsize
\textbf{Felicitas L\"offler*} \orcidicon{0000-0001-6423-7427} $^1$ 
\and \normalsize \textbf{Valentin Wesp} \orcidicon{https://orcid.org/0000-0002-8601-6032} $^1$ 
\and \normalsize \textbf{Birgitta K\"onig-Ries} \orcidicon{0000-0002-2382-9722}$^1$ $^2$ $^3$ 
\and \normalsize \textbf{Friederike Klan} \orcidicon{https://orcid.org/0000-0002-1856-7334} $^2$ $^4$}

\date{%
\normalsize
    $^1$Heinz-Nixdorf Chair for Distributed Information Systems, \\
    Department of Mathematics and Computer Science, \\
    Friedrich Schiller University Jena, Jena, Germany\bigskip \\%
    $^2$Michael-Stifel-Center for Data-Driven and Simulation Science, Jena, Germany\bigskip \\%
    $^3$German Center for Integrative Biodiversity Research (iDiv)\\
     Halle-Jena-Leipzig, Germany \bigskip\\%
    $^4$Citizen Science Group, DLR-Institute of Data Science\\
     German Aerospace Center, Jena, Germany\\[2ex]%
    \emph{\today}
}

\maketitle

* \emph{felicitas.loeffler@uni-jena.de}

\vspace{0.2in}
\begin{abstract}
\noindent The increasing amount of publicly available research data provides the opportunity to
link and integrate data in order to create and prove novel hypotheses, to repeat
experiments or to compare recent data to data collected at a different time or place.
However, recent studies have shown that retrieving relevant data for data reuse is a
time-consuming task in daily research practice.

In this study, we explore what hampers dataset retrieval in biodiversity research, a field that produces a large amount of heterogeneous data. We analyze the primary source in dataset search - metadata - and determine if they reflect scholarly search interests. We examine if metadata standards provide elements  corresponding to search interests, we inspect if selected data repositories use metadata standards  representing scholarly interests, and we determine how many fields of the metadata standards used are filled. To determine search interests in biodiversity research, we gathered 169 questions that researchers aimed to answer with the help of retrieved data, identified biological entities and grouped them into 13 categories. The categories were evaluated with nine biodiversity scholars who assigned one of the  types to  pre-labeled biological entities in the questions.

Our findings indicate that  environments, materials and chemicals, species, biological and chemical processes, locations, data parameters and data types are important search interests in biodiversity research.
The comparison with existing metadata standards shows that domain-specific standards cover search interests quite well, whereas general standards do not explicitly contain elements that reflect search interests. We inspect metadata from five large data repositories. Our results confirm  that metadata  currently poorly reflect search interests in biodiversity research. 
From these findings, we derive recommendations for researchers and data repositories how to bridge the gap between search interest and metadata provided.
\end{abstract}

\medskip


\section*{Introduction}

Scientific progress in biodiversity research, a field dealing with the diversity of life on earth - the variety of species, genetic diversity, diversity of functions, interactions and ecosystems \cite{idiv}, is increasingly achieved by the integration and analysis of heterogeneous datasets \cite{GBIF-SR2018,culina2018}. Therefore, locating and finding proper data for synthesis is a key challenge in daily research practice. Datasets can differ in format and size. Interesting data is often scattered across various repositories focusing on different domains. In a survey conducted by the Research Data Alliance (RDA) Data Discovery Group \cite{RDA2018}, 35\% of the 98 participating repositories stated that they host data from Life Science and 34\% indicated they cover Earth Science. All of these are potentially of interest to biodiversity researchers.

However, the offered search services at public data providers do not seem to support scholars effectively. A study by Kacprzak et al. \cite{Kacprzak2018} reports that 40\% of the users, who had sent data search requests to two open data portals, said, that they could not find the data they were interested in and thus directly requested the data from the repository manager. In several studies, ecologists report on the difficulties they had when looking for suitable datasets to reuse  \cite{Parker2016}  \cite{Ramakers2018} \cite{culina2018}. Scholars from research projects we are involved in also complain that data discovery is a time-consuming task. They have to search in a variety of data repositories with several different search terms to find data about species, habitats, or processes. Thus, there is a high demand for new techniques and methods to better support scholars in finding relevant data.

In this study, we explore what hampers data set retrieval in biodiversity research. We analyze two building blocks in retrieval systems: \emph{information needs (user queries)} and underlying \emph{data}. We want to find out how large the  gap is between scholarly search interests and provided data. In order to identify scholarly search interests, we analyzed \emph{user questions}. In contrast to user queries, which are usually formulated in a few keywords, questions represent a search context, a more comprehensive information need. Characteristic terms or phrases in these textual resources can be labeled and classified to identify biological entities \cite{Kilicoglu2018,BioASQ2017}. \emph{Scientific data} are not easily accessible by classical text retrieval mechanisms as they were mainly developed for unstructured textual resources. Thus, effective data retrieval heavily relies on the availability of proper \emph{metadata} (structured information about the data) describing available datasets in a way that enables their \emph{Findability}, one principle to ensure FAIR data \cite{FAIRPrinciples}. A survey conducted by the Research Data Alliance (RDA) Data Discovery Group points out that 58\% of the 98 participating data repositories index all metadata and partial metadata (52\%), and only 33\% integrate data dictionaries or variables \cite{RDA2018}.

We argue that \emph{Findability} at least partially depends on how well metadata reflect scholarly information needs. Therefore, we propose the following layered approach:\bigskip\\
\noindent
(A) At first, we identified main entity types (categories) that are important in biodiversity research. We collected $169$ questions provided by 73 scholars of three large and very diverse biodiversity projects in Germany, namely \emph{AquaDiva} \cite{AquaDiva}, GFBio - The German Federation for Biological Data \cite{GFBio} and iDiv - The German Research Center for Integrative Biodiversity Research \cite{idiv}. Two authors of this publication labeled and grouped all noun entities into $13$ categories (entity types), which were identified in several discussion rounds. Finally, all proposed categories were evaluated with biodiversity scholars in an online survey. The scholars assigned the proposed categories to important phrases and terms in the questions (Section ``\nameref{sec:questions}'').\\ 

\noindent
(B) Most data providers use keyword-based search engines returning data sets that exactly match  keywords entered by a user \cite{RDA2018}. In  dataset search, the main source are metadata that contain  structured entries on measurements, data parameters or species observed rather than  textual descriptions. It depends on the metadata schema used how sparse or rich the description turns out to be and which facets are provided for filtering. Therefore, we inspected common metadata standards in the Life Sciences and analyzed, to which extent their metadata schemes cover the identified information categories (Section ``\nameref{sec:metadata}'').\\

\noindent
(C) There are several data repositories that take and archive scientific data for biodiversity research. According to \emph{Nature's} list of recommended data repositories \cite{NatureRepoList}, repositories such as \emph{Dryad} \cite{Dryad}, \emph{Zenodo} \cite{Zenodo} or \emph{Figshare} \cite{Figshare} are generalist repositories and can handle different types of data. Data repositories such as \emph{Pangaea} \cite{Pangaea} (environmental data) or \emph{GBIF} \cite{GBIF} (taxonomic data) are domain specific and only take data of a specific format. We harvested and parsed all publicly available metadata from these repositories and analyzed, if they utilize metadata schemes with elements reflecting search interests. For \emph{GBIF}, we concentrated on  datasets only, as  individual occurrence records are not available in the metadata API. We explored how many fields of the respective schemas are actually used and filled (Section ``\nameref{sec:dataRepo}''). \\

\noindent
(D) Finally, we discuss the results and outline how to consider and address user interests in metadata (Section ``\nameref{sec:discussion}''). \\

\noindent
In order to foster reproducibility,  questions, scripts, results, and the parsed metadata are publicly available: 
\begin{small}\url{https://github.com/fusion-jena/QuestionsMetadataBiodiv}\end{small} \\

\noindent
The structure of the paper is as follows: The first part ``\nameref{sec:definition}'' focuses on the clarification of various terms. This is followed by sections that explain basics in Information Retrieval (``\nameref{sec:background}'') and ``\nameref{sec:relWork}''. The fourth section ``\nameref{sec:objectives}'' gives an overview of our research idea. The following four sections contain the individual research contributions described above. Each of these  sections describes the respective methodology and results. Finally, section ``\nameref{sec:conclusion}'' summarizes our findings.
\section*{Definitions}
\label{sec:definition}

Since dataset retrieval is a yet largely unexplored research field \cite{Chapman2019}, few definitions exist describing what it comprises and how it can be characterized. Here, we briefly introduce an existing definition and add our own definition from the Life Sciences' perspective.

Chapman et al \cite{Chapman2019} define a dataset as ``A collection of related observations organized and formatted for a particular purpose''. They further characterize a dataset search as an application that ``involves the discovery, exploration, and return of datasets to an end user.'' They distinguish between two types: (a) a basic search in order to retrieve individual datasets in data portals and (b) a constructive search where scholars create a new dataset out of various input datasets in order to analyze relationships and different influences for a specific purpose.

From our perspective, this definition of a dataset is a bit too restricted. All kinds of scientific data such as experimental data, observations, environmental and genome data, simulations and computations can be considered as datasets. We therefore extend the definition of Chapman et al \cite{Chapman2019} as follows:

\begin{mydef}
A \textbf{dataset} is a collection of scientific data including primary data and metadata organized and formatted for a particular purpose.
\end{mydef}

We agree with Chapman et al.'s definition of dataset search. We use \emph{Dataset Search} and \emph{Dataset Retrieval} synonymously and define it as follows:

\begin{mydef}
\textbf{Dataset Retrieval} comprises the search process, the ranking and return of scientific datasets.
\end{mydef}

Unger et al. \cite{Unger2014} introduced three dimensions to take into account in Question Answering namely the \emph{User} and \emph{Data} perspective as well as the \emph{Complexity} of a task. We argue that these dimensions can also be applied  in dataset retrieval.

\subsection*{User Perspective}
In conventional retrieval systems users' search interests are represented as a few keywords that are sent to the system as a search query. Keywords are usually embedded in a search context that can be expressed in a full sentence or a question.


%

In order to understand what users are looking for, a semantic analysis is needed. \emph{Information Extraction} is a technique from text mining that identifies main topics (also called entity types) occurring in unstructured text \cite{Jurafsky2008}. Noun entities are extracted and categorized based on rules. Common, domain-independent entity types are for instance \emph{Person}, \emph{Location}, and \emph{Time}. When it comes to specific domains, additional entity types corresponding to core user interests need to be taken into consideration. In bio-medicine, according to \cite{bioCADDIE}, the main topics are data type, disease type, biological process and organism. In new research fields such as biodiversity research  these main entity types still need to be identified in order  to get insights into users' information needs and to be able to later adapt systems to user requirements.

\subsection*{Data Perspective}

From the data perspective, a dataset search can be classified into two types based on the source of data: \emph{primary data} and \emph{metadata}.

\begin{mydef}
\textbf{Primary data} are scientific raw data. They are the result of scientific experiments, observations, or simulations and vary in type, format, and size.
\end{mydef}

\begin{mydef}
\textbf{Metadata} are structured, descriptive information of primary data and answer the W-questions: \emph{What?} has been measured by Whom?, When?, Where? and Why?. Metadata are created for different purposes such as search, classification, or knowledge derivation.
\end{mydef}


Dataset retrieval approaches focussing on primary data as source data have to deal with different data formats such as tabular data, images, sound files, or genome data. This requires specific query languages such as QUIS \cite{Chamanara2017} to overcome the ensuing heterogeneity and is out of scope of this paper. Here,  
we solely focus on dataset retrieval approaches that use metadata as input for search. A variety of metadata standards in the Life Sciences are introduced in Section ``\nameref{sec:metadata}''.

\subsection*{Complexity}

Scholarly search interests are as heterogeneous as data are. Information needs can range from specific questions where users expect datasets to contain the complete answer to broader questions that are answered partially only by datasets. Furthermore, users construct new datasets out of various input datasets. Unger et al. \cite{Unger2014} characterize the complexity in retrieval tasks along four dimensions: \emph{Semantic complexity} describes how complex, vague, and ambiguous a question is formulated and if heterogeneous data have to be retrieved. \emph{Answer locality} denotes if the answer is completely contained in one dataset or if parts of various datasets need to be composed or if no data can be found to answer the question. \emph{Derivability} describes if the answer contains explicit or implicit information. The same applies for the question. If broad or vague terms appear in the question or answer, additional sources have to be integrated to enrich both, question and/or answer. \emph{Semantic tractability} denotes if the natural language question can be transformed into a formal query.

In this work, we do not further explore the complexity of questions. We focus on the analysis of user interests and metadata, only.
\section*{Background}
\label{sec:background}
This section provides  background information on which parts are involved in a search process, how the system returns a result based on a user's query and what evaluation methods and metrics exist in Information Retrieval.

\subsection*{The Retrieval Process}
\begin{figure}[t]
\centering
\includegraphics[width=0.8\textwidth]{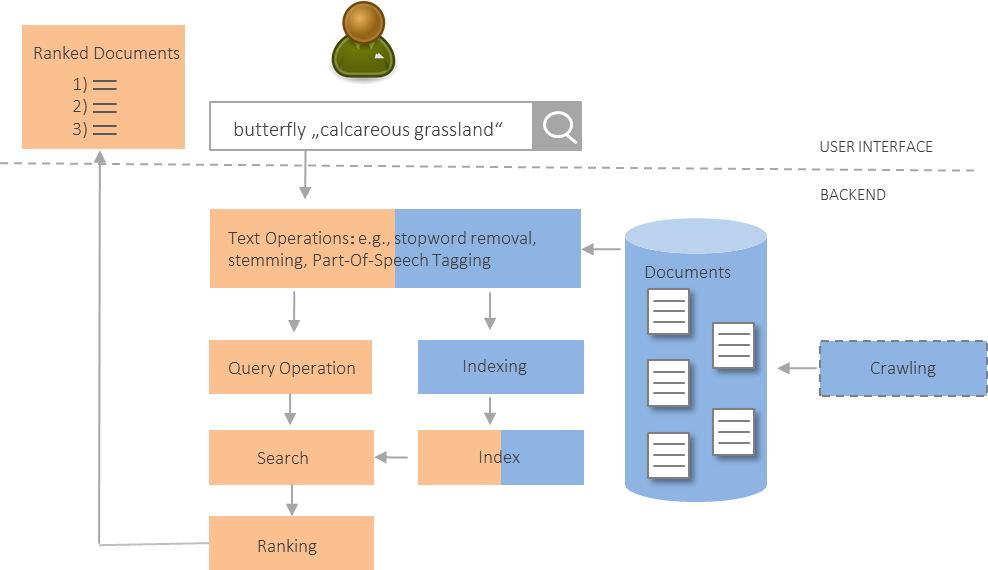}\hfill
\caption{The architecture of an Information Retrieval system based on \cite{MIR2011}: An optional \emph{Crawling Process} (blue, dashed line) gathers documents or web pages. In the \emph{Indexing Process} (blue) the documents are pre-processed before an index can be established. The \emph{Retrieval Process} (orange) comprises the transformation of a user query into a format the search engine understands before the actual search and ranking takes place. Finally, users receive a ranked list of documents that match their query.} \label{fig:retrievalProcess}

\end{figure}

A retrieval system consists of a collection of documents (a \emph{corpus}) and a user's information needs that are described with a few keywords (\emph{query}). The main aim of the retrieval process is to return a ranked list of documents that match a user's query. The architecture of a retrieval system is depicted in Figure (\ref{fig:retrievalProcess}): If the document corpus is not given, an optional \emph{Crawling Process} has to be run beforehand to retrieve and collect documents
\cite{MIR2011}. The \emph{Indexing Process} comprises pre-processing steps such as stopword removal, stemming, and spell checks  important to clean documents from unnecessary information and to analyze only those terms that truly represent the content of a document. Afterwards, the system counts word frequencies within a document and across all documents. The result is an inverted index. Similar to a book index, this is a list of terms together with the number of occurrences of each term in each document and across all documents.  These statistics, generated regularly in background processes, form the basis for a fast access to the documents at search time. The actual search takes place in the \emph{Retrieval and Ranking Process} whenever a user sends a query to the system and results in a ranked result set being returned to the user.

Based on the underlying \emph{Retrieval Model}, different ranking functions have been developed to produce a score for the documents with respect to the query. Top-scored documents are returned first. In larger corpora, paging functions allow a subsequent retrieval of further  documents. Classical retrieval models are for instance: the \emph{Boolean Model} \cite{Manning2008} where only documents are returned that exactly match a query. In this model all documents in the retrieved set are equally relevant and therefore it is not considered as a ranking algorithm. It is often used in search engines in combination with further retrieval models such as the \emph{Vector Space Model} \cite{Manning2008}. Here, documents are represented by vectors that consist of term weights. The similarity of documents and queries is determined by computing the distance between the vectors. \emph{Probabilistic Models} \cite{Manning2008} are based on computations of the probability of  a document belonging to the relevant set. For languages where word boundaries are not given, e.g., in Eastern Asian Languages, \emph{Language Models}\cite{Jurafsky2000} have to be applied to get a mathematical representation of the documents. The system analyzes the text documents by means of character-based sliding windows (\emph{n-grams}) to determine word boundaries and compute statistics. All these classical retrieval models are keyword-based. Thus, retrieval systems only return documents that exactly match the  user query.

\subsection*{Evaluation in Information Retrieval}

When setting up a retrieval system, various design decisions influencing different parts of the system have to be made. Examples of such decisions are whether to stem terms in the pre-processing phase or which terms to include in the stopword list. 

Numerous evaluation measures have been developed to determine the \emph{effectiveness} of the systems, i.e.,  the accuracy of the result returned by a given retrieval algorithm.
 For this purpose, a test collection is required that consists of three things \cite{Manning2008}: (1) a corpus of documents, (2) representative information needs expressed as queries, (3) a set of relevance judgments provided by human judges containing assessments of the relevance of a document for given queries. If judgments are available for the entire corpus they serve as baseline (``gold standard'') and can be used to determine how many relevant documents a search system finds for a specific topic.

User queries should be representative for the target domain. Queries are either obtained from query logs of a similar application or domain users are asked to provide example queries \cite{Croft2009}. The number of example questions influences the evaluation result.  TREC (Text REtrieval Conference) is a long-running, very influential annual Information Retrieval competition that considers different retrieval issues in a number of \emph{Tracks}, e.g., Genomics Track or Medical Track (\url{https://trec.nist.gov/}). Various TREC experiments have shown that the number of queries used for the evaluation matters more than the number of documents judged per query \cite{Croft2009}. Therefore, TREC experiments usually consist of around 150 queries (or so-called ``topics'') per track.

Common evaluation metrics with respect to effectiveness are \emph{Precision and Recall (PR)}, \emph{F-Measure} and \emph{Mean Average Precision (MAP)} \cite{Manning2008}. Precision denotes which fraction of the documents in the result set is relevant for a query, whereas  recall describes which fraction of  relevant documents was successfully retrieved. Both metrics are based on binary judgments, i.e.,  raters can only determine, if a document is relevant or non-relevant. The F-Measure is the harmonic mean of Precision and Recall. 

Precision and Recall can only be used, when a gold standard is provided containing the total number of documents in a corpus that are relevant for a query. However, in applied domains where corpora are established specifically for a particular research field, gold standards are usually not available. Therefore, with the recall being unknown,  \emph{MAP} only requires to get ratings for the TopN-ranked documents to compute an average precision. The assumption here is that users are only interested in the first entries of a search result and usually do not navigate to the last page. The top-ranked documents get higher scores than the lower ranked ones \cite{Croft2009}. Another metric proposed by  J{\"a}rvelin and Kek{\"a}l{\"a}inen\cite{Jarvelin2002} is the \emph{Discounted Cumulated Gain (DCG)}, a metric that uses a Likert-scale as rating scheme and allows non-binary ratings. All entries of the scheme should be equally distributed. However, DCG does not penalize for wrong results but only increases the scores of top-ranked documents.
Other evaluation criteria concentrate on the \emph{efficiency} (e.g., the time, memory and disk space required by the algorithm to produce the ranking \cite{Croft2009}), \emph{user satisfaction} on the provided result set and \emph{visualization} \cite{MIR_UIViz2011}.

\begin{flushleft}
The missing aspect in evaluation approaches in Information Retrieval is the analysis of the underlying documents. The data source in classical Information Retrieval systems is unstructured text whereas Dataset Retrieval is based on structured metadata files. Hence, retrieval success depends on the metadata format used, the experience of the curator, and the willingness of the individual scholar to describe  data properly and thoroughly. This structured information could be used in the search index. For instance, if researchers provide information such as taxon, length, location or experimental method in the metadata explicitly, a search application could offer a search over a specific metadata field. Thus, we argue that there is a need to analyze the given metadata and to quantify the gap between a scholar's actual search interests and the metadata  primarily used in search applications.
\end{flushleft}

\section*{Related Work}
\label{sec:relWork}
This section focuses on approaches that analyze, characterize and enhance dataset search.
We discuss studies identifying users' information needs and introduce existing question corpora. In a second part, we describe approaches that aim at improving dataset search.
%


\subsection*{User Interests}
In order to understand user behavior in search, query logs or question corpora are valid sources. Kacprazal et al \cite{Kacprzak2018} provide a comprehensive log analysis of three government open data portals from the United Kingdom (UK), Canada, and Australia and one open data portal with national statistics from the UK.
2.2 million queries  from  logs provided by  the data portals (internal queries) and 1.1 million queries  issued to external web search engines (external queries) were analyzed. Two authors manually inspected a sample set of 665 questions and determined the main query topics. Most queries were assigned to Business and Economy (20$\%$ internal queries, 10$\%$ external queries) and Society (14.7$\%$ internal queries, 18$\%$ external queries). Besides query logs,  Kacprazal et al \cite{Kacprzak2018} also  explicit requests by users for data via a form on the website. Here, users provided title and description which allowed the authors to perform a deeper thematic analysis on 200 manually selected data requests. It revealed that geospatial (77.5$\%$) and temporal (44$\%$) information occurred most, often together with a specific granularity (24.5$\%$), e.g., ``hourly weather and solar data set'' or ``prescription data per hospital''. Users were also asked why they had requested data explicitly, and more than 40$\%$ indicated that they were not able to find relevant data via the provided search.

In the Life Sciences, Dognan et al \cite{Dogan2009} inspected one month of log data with more than 58 million user queries from PubMed \cite{PubMed}, a platform providing biomedical literature. They randomly selected 10,000 queries for a semantic analysis. Seven annotators categorized the queries along 16 given categories. They distinguished between bibliographic queries (44$\%$) containing information such as journal name, author name, or article title and non-bibliographic queries with domain specific categories. The most frequent category over all questions was ``Author Name'' (36$\%$) followed by ``Disorder'' (20$\%$) comprising diseases, abnormalities, dysfunctions etc., and ``Gene/ Protein'' (19 $\%$). Further main topics were abbreviations (mostly from genes/ proteins) and chemicals/drugs.

A large study on user needs in biodiversity research have been conducted in the GBIF community in 2009 \cite{Faith2013,Arino2013}. The aim was to determine what GBIF users need in terms of primary data and to identify data gaps in the current data landscape at that time. More than 700 participants from 77 countries took part in the survey. It revealed that scholars used retrieved primary data for analyzing species diversity, taxonomy, and life histories/ phenology. That mainly required ``taxon names, occurrence  data and  descriptive  data  about the  species'' \cite{Arino2013}. As biodiversity is a rapidly changing research field, the authors recommend to repeat content need assessments in frequent intervals \cite{Faith2013}. \\

\noindent
Apart from query logs, question corpora are another source for identifying search interests. Usually, questions are collected from experts of a particular research field and important terms representing main information needs are labeled with categories or so-called \emph{entity types}. These manually generated annotations help  understanding what information users are interested in and  developing tools and services to either automatically extract these interests from text (Text Mining), to retrieve relevant data (Information Retrieval) or to provide an exact answer for that information need (Question Answering).

In the Life Sciences, question corpora for text retrieval have been mainly established  in the medical and biomedical domains. One of the largest corpora in medicine is the Consumer Health Corpus \cite{Kilicoglu2018}, a collection of email requests (67\%) received by the U.S. National Library of Medicine (NLM) customer service and search query logs (33\%) of MedlinePlus, a consumer-oriented NLM website for health information. The final corpus consists of 2614 questions and has been integrated into the Medical Question Answering Task at TREC 2017 LiveQA \cite{LiveQA2017}. Six trained domain experts were involved in the annotation tasks to manually label information. The experts had to indicate named entities, e.g., problem, anatomy or measurement and labeled question topics such as the cause of a disease or complications (longer term effects of a disease).

A common question corpus in biomedicine is the Genomics Track at TREC conferences \cite{GenomicsTrack}. The topics of the retrieval tasks are formulated as natural language questions and contain pre-labeled main categories, e.g., \emph{What [GENES] are involved in insect segmentation?}. A further large question corpus in biomedicine is the question corpus created for the BioASQ challenge \cite{BioASQ2017}, an annual challenge for researchers working on text mining, machine learning, information retrieval, and question answering. The tasks are split into three parts: (1) the extraction of main entities and their linkage with ontological concepts (semantic annotation), (2) the translation of natural language queries into RDF triples, and (3) the retrieval of the exact answer to a natural language query. The question corpus was created and annotated by a team of $10$ experts, selected with the goal to cover different ages and complementary expertise in the fields of medicine, biology, and bioinformatics \cite{BioASQTeam2013}. Each expert was asked to formulate $50$ questions in English that reflect ``real-life information needs''. However, the type of questions to be formulated was restricted, e.g., the experts were instructed to provide questions of certain types typically considered in question answering systems (yes/no, factoid, etc.). These restrictions are justified to a certain degree since they affect the applicability of the resulting corpus for evaluation purposes of question answering approaches. However, they have an impact on which questions are formulated and how. This will likely  lead to a bias in the question corpus.

Another question corpus in the biomedical domain is the benchmark developed for the 2016 bioCADDIE Dataset Retrieval Challenge \cite{bioCADDIE}. This benchmark was explicitly created for the retrieval of datasets based on metadata and includes 137 questions, 794,992 datasets gathered from different data portals in XML structure, and relevance judgments for 15 questions. Similar to the BioASQ challenge, domain experts got instructed on how to create questions. Based on templates, the question constructors formulated questions using the most desired entity types, namely data type, disease type, biological process, and organism.\\

\noindent
At present, to the best of our knowledge, there is neither a public log analysis nor a question corpus available for biodiversity research. In order to understand genuine user interests and to improve current dataset retrieval systems, unfiltered information needs are crucial.
Therefore, collecting current search interests from scholars is the first step in our top-down approach presented in Section ``\nameref{sec:objectives}''.

%
%


\subsection*{Dataset Search}
A study by the RDA Data Discovery Group points out \cite{RDA2018} that most data repositories offer search applications based on metadata and utilize one of the existing and widely spread search engines for data access, e.g., \emph{Apache Solr} (\url{http://lucene.apache.org/solr/}) or \emph{elasticsearch} (\url{https://www.elastic.co/products/elasticsearch}). Large data repositories such as \emph{GBIF} \cite{GBIF_Search}, \emph{PANGAEA} \cite{Pangaea_Search} or \emph{Zenodo} \cite{Zenodo_Search} also use \emph{elasticsearch} and offer public search services. \emph{Apache Solr} and \emph{elasticsearch} are both keyword-based and return datasets that exactly match a user's entered query terms. If the desired information need is not explicitly mentioned in the metadata, the search will fail.

In recent years, a variety of approaches have emerged to improve dataset search. A common approach is to annotate metadata with entities from \emph{schema.org} (\url{https://schema.org}). Favored by Google \cite{GoogleSchema} and the RDA Discovery Task Group \cite{RDASchema}, the idea is to add descriptive information to structured data such as XML or HTML in order to increase findability and interoperability. These additional attributes help search engines to better disambiguate  terms occuring in text. For example, Jaguar could be a car, an animal or an operating system. By means of \emph{schema.org} entities, data providers can define the context explicitly.
Numerous extensions for specific domains have been developed or are still in development, e.g., \emph{bioschemas.org} \cite{bioschema2018} for the Life Sciences. Since Google launched its beta version of a dataset search in Fall 2018 (\url{https://toolbox.google.com/datasetsearch}), \emph{schema.org} entities got more and more attention. Hence, data centers such as \emph{PANGAEA} \cite{Pangaea} or \emph{Figshare} \cite{Figshare} are increasingly incorporating \emph{schema.org} entities in their dataset search.

Other approaches favor an improved metadata schema. Pfaff et al \cite{EASE2017} introduce the Essential Annotation Schema for Ecology (EASE). The schema was primarily developed in workshops and intensive discussions with scholars and aims to support scientists in search tasks. The MIBBI project \cite{MIBBI2008} (now known as \emph{BioSharing} or \emph{FAIRSharing} portal - \url{https://fairsharing.org/}) also recognized that only improved metadata allow information seekers to retrieve relevant experimental data. They propose a harmonization of minimum information checklists in order to facilitate data reuse and to enhance data discovery across different domains. Checklist developers are advised to consider ``'cross-domain' integrative activities''\cite{MIBBI2008} when creating and maintaining checklists. In addition, standards are supposed to contain information on formats (syntax), vocabularies and ontologies used.

The latter points to an increasing interest in semantic techniques that have emerged over the past decade. Vocabularies such as the Data Catalog Vocabulary (DCAT) \cite{DCAT} or the Vocabulary of Interlinked Datasets (VoID) \cite{VOID} aim to describe datasets semantically in RDF \cite{RDFSchema} or OWL \cite{OWL} format based on subject, predicate, and object triples. Fully semantic approaches such as BioFED \cite{BioFED2017} offer a single-point-of-access to 130 SPARQL endpoints in the Life Sciences. They integrate a variety of heterogeneous biomedical ontologies and knowledge bases. Each data source is described by VoID descriptors that facilitate federated SPARQL query processing. The user interface permits simple and complex SPARQL queries and provides support in creating federated SPARQL queries. The result set contains provenance information, i.e., where the answer has been found, ``the number of triples returned and the retrieval time''\cite{BioFED2017}. However, improvements in the user interface still remain necessary. As BioFED is mainly focused on a linked data approach, it requires  all data sources to be stored in semantic formats and users  to have at least basic SPARQL knowledge. In contrast, Kunze and Auer \cite{KunzeAuer2013} consider the search process in their search over RDF datasets as an exploratory task based on semantic facets. Instead of SPARQL queries or keyword-based user interfaces, they provide parameters for filtering. This allows an unambiguous search and returns relevant datasets that match the provided filter parameters.

Other federated approaches outside semantic techniques attempt to align heterogeneous data sources in one search index. That allows the use of conventional search engines and keyword-based user interfaces: DataONE is a project aiming to provide access to earth and environmental data provided by multiple member repositories \cite{DataONE2012}. Participating groups can provide data in different metadata formats such as EML, DataCite or FGDC \cite{DataOneIndexer}. DataONE is currently working on quantifying FAIR \cite{DataOneFairCheck}. Their findability check determines if specific metadata items such as title, abstract or publication date are present. For title and abstract, they additionally check the length and content. Based on these criteria, they evaluated their data and found out that concerning \emph{Findability} around 75\% of the available metadata fulfilled the self-created criteria.
The German Federation for Biological Data (GFBio) \cite{GFBio2014} is a national infrastructure for research data management in the green Life Sciences and provides a search over more than six million heterogeneous datasets from environmental archives and collection data centers. It was extended to a semantic search \cite{HoneyBee2017} that allows a search over scientific names, common names, or other synonyms. These related terms are obtained from GFBio's Terminology Service \cite{GFBioTS2016} and are added in the background to a user's query.

As described above, numerous approaches have been proposed and developed to improve dataset search. However, what is lacking is a comprehensive analysis on what exactly needs to be improved and how large the actual gap is between user requirements and given metadata.

\section*{Objectives}
\label{sec:objectives}

Current retrieval evaluation methods are basically focused on improving retrieval algorithms and ranking. Therefore, question corpora and documents are taken as given and are not questioned. However, if the underlying data do not contain the information users are looking for, the best retrieval algorithm will fail. We argue, in  dataset search, metadata, the basic source for dataset applications, need to be adapted to match users' information needs. 
\\
\noindent
We want to find out how large the gap in biodiversity research is between actual user needs and provided metadata  and how to overcome this obstacle. Thus, the following analysis aims to explore:

\begin{itemize}
	\item What are genuine user interests in biodiversity research?
	\item Do existing metadata standards reflect information needs of biodiversity scholars?
	\item Are metadata standards utilized by data repositories  useful for data discovery? How many metadata fields are filled?
	\item Do common metadata fields contain useful information?
\end{itemize}

\noindent
We take a top-down approach starting from  scholars' search interests, then looking at metadata standards and finally inspecting the metadata provided in selected data repositories. \\

\noindent
(A) First, we generate an annotated question corpus for the biodiversity domain: We gather questions from scholars, explore the questions and identify information categories. In an online evaluation, domain experts assign these categories to terms and phrases of the questions (Section ``\nameref{sec:questions}''). \\

\noindent
(B) We inspect different metadata standards in the Life Sciences and compare the metadata elements to the identified search categories from (A) (Section ``\nameref{sec:metadata}'').\\

\noindent
(C) We analyze the application programming interfaces (APIs) of selected data repositories to figure out what metadata standards are used and how many elements of a metadata schema are utilized for data description (Section ``\nameref{sec:dataRepo}'').\\

\noindent
(D) We discuss how to bridge the gap between  users' search interests and metadata. We propose an approach to overcome the current obstacles in dataset search (Section ``\nameref{sec:discussion}'').

\section*{A - Information Needs in the Biodiversity Domain}
\label{sec:questions}

Question corpora are common sources for getting an impression what users are interested in in a particular domain. Therefore, we asked biodiversity scholars to provide questions that are specific for their research. We analyzed the questions and identified search topics that represent scholarly information needs in this domain.

\subsection*{Methodology}
The following subsection describes the methodology in detail, divided into four paragraphs.

\paragraph*{Questions:}

We gathered questions in three large biodiversity projects, namely \emph{CRC AquaDiva} \cite{AquaDiva}, \emph{GFBio} \cite{GFBio} and \emph{iDiv} \cite{idiv}. We explicitly requested fully expressed questions to capture the keywords in their search context. These projects vary widely in their overall setting, the scientists and disciplines involved and their main research focus. Together, they provide a good and rather broad sample of current biodiversity research topics.
In total, 73
scholars with various research backgrounds in biology (e.g., ecology, bio-geochemistry, zoology and botany) and related fields (e.g., hydro-geology) provided 184 questions. This number is comparable to related question corpora in Information Retrieval (e.g., bioCADDIE \cite{bioCADDIE}) which typically consist of around  100 – 150 questions. The scholars were asked to provide up to five questions from their research background. Questions varied with respect to granularity. The corpus contains specific questions, such as \emph{List all datasets with organisms in water samples!} or questions with a broader scope, e.g., \emph{Does agriculture influence the groundwater?}. We published the questionnaires that were handed out in AquaDiva and iDiv as supplementary material in our repository. In the GFBio project, questions were gathered via email and from an internal search evaluation. All questions were inspected by the authors with respect to comprehensibility. We discarded questions which were not fully understandable (e.g., missing verb, misleading grammatical structures) but left clear phrases in the corpus that were not fully expressed as a question. If scholars provided several questions, they were treated individually even if terms referred to previous questions, e.g., \emph{Do have earthworm burrows (biopores) an impact on infiltration and transport processes during rainfall events?} and \emph{Are the surface properties influencing those processes?}. In this case, no further adaption towards comprehensibility has been made. The questions were also not corrected with respect to grammar and spelling since changing the grammar could lead to an altered statement. We did not want to loose the original question statement. In some questions, abbreviations occurred without explanations. In these cases, we left the questions as they are and did not provide full terms, since these abbreviations can have various meanings in different biological fields. It was up to the domain experts to either look them up or to leave the term out.
After the cleaning, the final corpus consists of 169 questions and is publicly available: \url{https://github.com/fusion-jena/QuestionsMetadataBiodiv/tree/master/questions}.

%
%
%

\paragraph*{Categories:}

Boundaries of semantic categories are domain-dependent and fuzzy. However, in search, categories support users in finding relevant information more easily and should be valid across various research backgrounds.
In a first round, two authors of this work analyzed the collected questions manually. Both have a research background in computer science and strong knowledge in scientific data management, in particular for biodiversity research. The corpus was split up and each of them inspected around 50\% of it and assigned broad categories independently of the other one. Afterwards, this first classification was discussed in several sessions. This resulted in 13 categories. The naming was adapted to domain-specific denotations and ontologies. Furthermore, the categories were compared to EASE \cite{EASE2017}, a metadata schema which was primarily developed for an improved dataset retrieval in the field of ecology. This comparison revealed that there is an overlap with EASE but that we discovered further relevant categories \cite{Loeffler2017}.
The final categories are:
\begin{enumerate}
\item ORGANISM comprises all individual life forms including plants, fungi, bacteria, animals and microorganisms.
\item All species live in certain local and global ENVIRONMENTS such as habitats, ecosystems (e.g., below 4000 m, ground water, city) and
\item have certain characteristics (traits, phenotypes) that are summarized with QUALITY \& PHENOTYPE, e.g., length, growth rate, reproduction rate, traits.
\item Biological, chemical and physical PROCESSES are re-occurring and transform materials or organisms due to chemical reactions or other influencing factors.
\item EVENTS are processes that appear only once at a specific time, such as environmental disasters, e.g., Deepwater Horizon oil spill, Tree of the Year 2016.
\item Chemical compounds, rocks, sand and sediments can be grouped as MATERIALS \& SUBSTANCES.
\item ANATOMY comprises the structure of organisms, e.g., body or plant parts, organs, cells, and genes.
\item METHOD describes all operations and experiments that have to be conducted to lead to a certain result, e.g., lidar measurements, observation, remote sensing.
\item Outcomes of research methods are delivered in DATA TYPE, e.g., DNA data or sequence data is the result of genome sequencing, lidar data is the result of lidar measurements (active remote sensing).
\item All kinds of geographic information is summarized with LOCATION, e.g., Germany, Hainich, Atlantic Ocean, and
\item temporal data including date, date times, and geological eras are described by TIME, e.g., current, over time, triassic.
\item PERSON \& ORGANIZATION are either projects or authors of data.
\item As reflected in the search questions, scholars in biodiversity are highly interested in HUMAN INTERVENTION on landscape and environment, e.g., fishery, agriculture, and land use.
\end{enumerate}

\noindent
For the evaluation with domain experts we added two more categories, namely OTHER and NONE. The first permits to define an own category, if none of the given ones is appropriate. NONE applies, if the term is not relevant, or if the domain expert does not know the term or if the phrase is too fuzzy and can not be classified into one category.

\paragraph*{Annotation:}

An annotation process usually has two steps: (1) the identification of terms based on annotation rules and (2) the assignment of an appropriate category in a given context. Usually, an annotator -  a domain expert -  who is trained in the annotation guidelines, carries out both tasks. However, we argue that training is somewhat biased and influences annotators in their classification decision. This is an obstacle in search where an intuitive feedback for category assignment is required. Hence, we split up the annotation process. Two scholars, who collected the questions and who are familiar with the guidelines conducted the identification, whereas domain experts only received short instructions and assigned categories. Our annotation guidelines, needed to identify phrases and terms (artifacts) to label, are available as supplementary material in our repository.

\paragraph*{Annotators and Annotation Process:}

Nine domain experts (8 Postdocs, 1 Project Manager) with expertise in various biological and environmental sciences participated in the classification task. All of them have experience in ecology but in addition, each of them has individual research competence in fields such as bio-geography, zoology, evolutionary biology, botany, medicine, physiology, or biochemistry.

\begin{figure*}[t]
\centering
\includegraphics[width=0.99\textwidth]{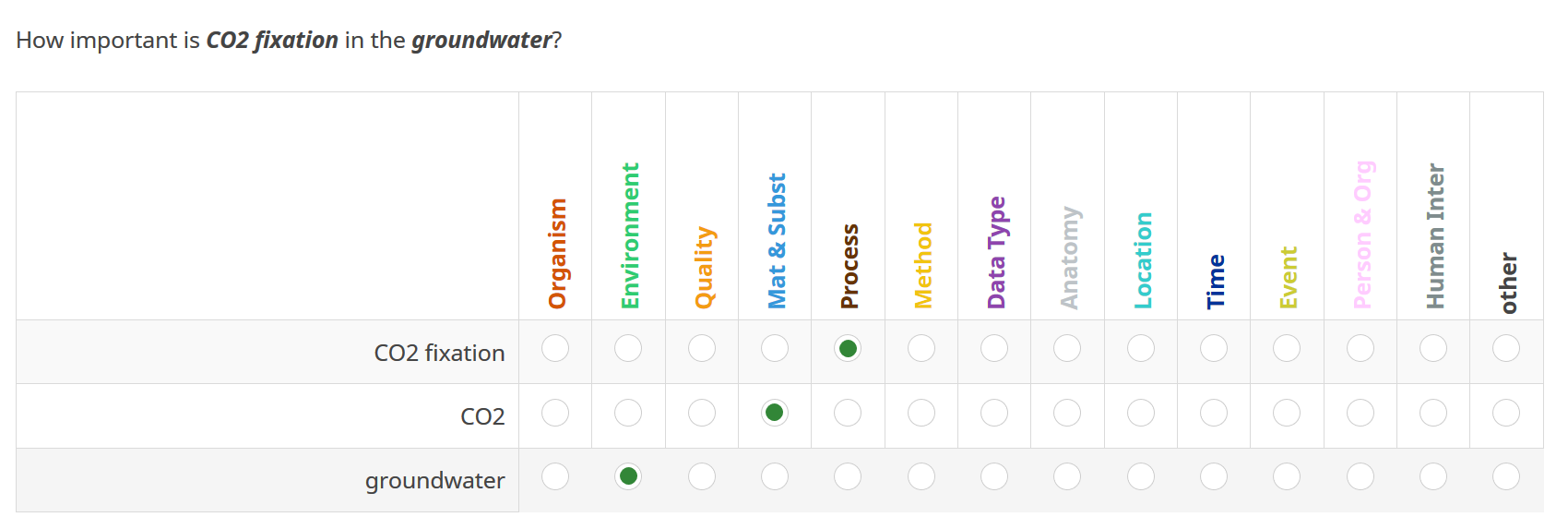}
\caption{Excerpt of the survey that was set up for the classification task. The annotators were told to assign only one category per given artifact. If an artifact is a compound noun, the nested entities such as adjectives or second nouns that further describe the term were provided for tagging as well.}
\label{fig:limesurvey}
\vspace*{-.5cm}
\end{figure*}

For the category assignment, all scholars received a link to an online survey with explanations of the categories (including examples) and short instructions on how to classify the artifacts. A screenshot of the survey is presented in Figure \ref{fig:limesurvey}. The purpose of this evaluation was also explained to them (improvement of data set retrieval systems). Multi-labeling was not allowed;  only one category was permitted per artifact. Should there be no proper category, they were advised to select OTHER and if possible to provide an alternative category. If they did not know a term or phrase, they could decide either to look it up or to omit it. The latter also applied,if they considered a phrase or term to be not relevant or too complicated and fuzzy. As we wanted to obtain intuitive feedback, the experts were told not to spend too much time on the classification decision but to determine categories according to their knowledge and research perspective. The annotators also had the opportunity to skip an artifact. In this case the category NONE was applied. For each question, annotators had the opportunity to provide a comment.

We decided to use a combination of csv files, Python scripts and \emph{Limesurvey} to support the annotation process.  Details on this process can be found in the supplementary material in our repository.

\subsection*{Results}
\label{sec:resQ}

We analyzed the user responses to determine whether the identified information categories are comprehensive and representative for biodiversity research. We computed the inter-rater agreement per artifact to determine the category that best describes an artifact.

\begin{figure}[t]
\centering
\includegraphics[width=0.9\textwidth]{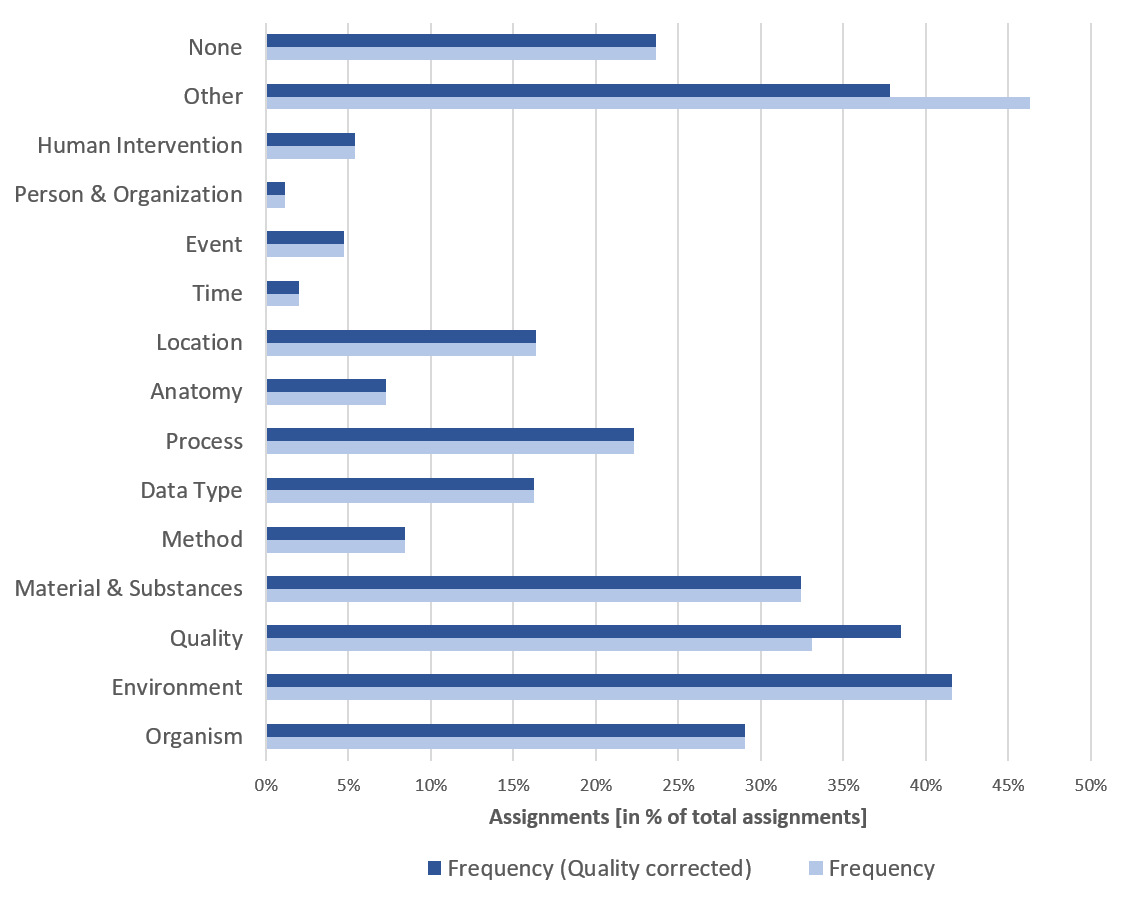}\hfill
\caption{The frequency of the categories and how often they were assigned to given phrases and terms, with and without QUALITY correction.} \label{fig:categoryfrequency}

\end{figure}

\paragraph{Representativeness of the Categories}
In order to verify completeness 
we determined the fraction of artifacts assigned to the category OTHER, i.e., if the experts deemed none of the given categories as appropriate. Figure~\ref{fig:categoryfrequency} depicts the frequency of information categories and how often they were selected by the domain experts. As it turned out, the category OTHER was selected by at least $1$ expert per artifact for $46\%$ of the phrases and terms and by at least $2$ experts for $24\%$. The fraction of phrases for which at least $3$ experts selected the category OTHER was $12\%$.
If at least two domain experts agree that there is no proper category for a given phrase, it is a strong indicator for a missing category or a misinterpretation. This is the case for $24\%$ out of all annotated artifacts. Hence, the coverage of the identified information categories is still high.


\noindent
However, there might be various reasons why none of the given categories fit:
(1) The phrase or term to be annotated was unknown to the annotator such as \emph{shed precipitation}. (2) Frequently, phrases that refer to
data attributes (e.g., \emph{soil moisture}, \emph{oxygen uptake rate} or \emph{amount of rain}) and which were supposed to be covered by the category QUALITY, were classified as OTHER. As alternative category, the annotators proposed ``Parameter'' or ``Variable''. When adding these ratings to the QUALITY category, the results for the OTHER category decreased to $37\%$/$13\%$/$4\%$.
That strongly indicates that renaming the QUALITY category or adding synonyms would increase comprehensibility significantly.
(3) The category OTHER was often chosen for terms used in questions with a broader scope in order to express expected results. However, since this is often vague, scholars tend to use generic terms such as \emph{signal}, \emph{pattern}, \emph{properties}, \emph{structure}, \emph{distribution}, \emph{driver} or \emph{diversity}. 
Hence, further discussions in the biodiversity research community are needed to define and classify these terms.



%

In addition, we wanted to know if there are categories that were not or rarely used by the annotators. This would indicate a low relevance for biodiversity research.  As depicted in Figure~\ref{fig:categoryfrequency}, the categories ENVIRONMENT, ORGANISM, MATERIAL \& SUBSTANCES, QUALITY, PROCESS, LOCATION and DATA TYPE have been selected most frequently (assigned to more than $15\%$ of the phrases). 
Information related to these categories seems to be essential for biodiversity research. Although there were categories that were rarely chosen (PERSON \& ORGANISATION and TIME), there was no category that was not used at all.

\paragraph{Consensus of the Categories}
In statistics, the consensus describes how much homogeneity exists in ratings among domain experts. We determined the inter-rater agreement and inter-rater reliability using Fleiss' Kappa ($\kappa$ statistics) \cite{Fleiss1971} and GWET's AC \cite{GWET2008}.
In general, the inter-rater reliability computes the observed agreement among raters ``and then adjusts the result by determining how much agreement could be expected from random chance''\cite{QuarfootLevine2016}. $\kappa$  values vary between $-1$ and $+1$, where values less than $0$ denote poorer than chance agreement and values greater than $0$ denote better than chance agreement. As suggested by Landis and Koch \cite{LandisKoch1977}, $\kappa$ values below $0.4$ indicate fair agreement beyond chance, values between $0.4$ and $0.6$ moderate agreement, values between $0.6$ and $0.8$ substantial agreement and values higher than $0.80$ indicate almost perfect agreement.
However, $\kappa$ statistics can lead to a paradox: When the distribution of the raters' scores is unbalanced, the correction for the chance agreement can result in negative $\kappa$ values even if the observed agreement is very high \cite{QuarfootLevine2016}. Since this is the opposite of what is expected, a new and more robust statistic has emerged, the GWET's AC \cite{GWET2008}. GWET's AC considers the response categories in the agreement by chance and the values can range from $0$ to $1$.

\begin{figure}[t]
\begin{adjustwidth}{-0.5in}{-0.5in}
\centering
\subfloat[Fleiss' Kappa values per category with and without QUALITY correction.\label{fig:fleisskappaQualityCorr}]{
\includegraphics[width=0.55\textwidth]{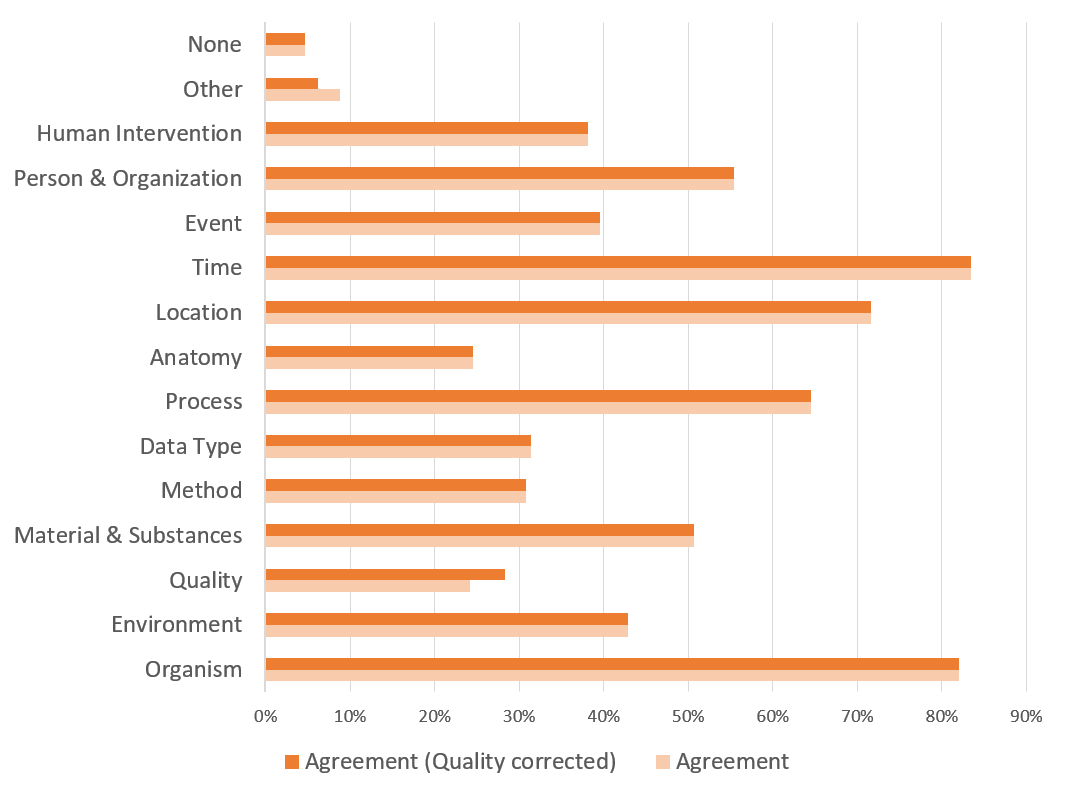}}\hfill
\subfloat[Fleiss' Kappa values per category for artifacts with one and two terms (with QUALITY correction).\label{fig:fleisskappaOneTwoTerms}] {\includegraphics[width=0.55\textwidth]{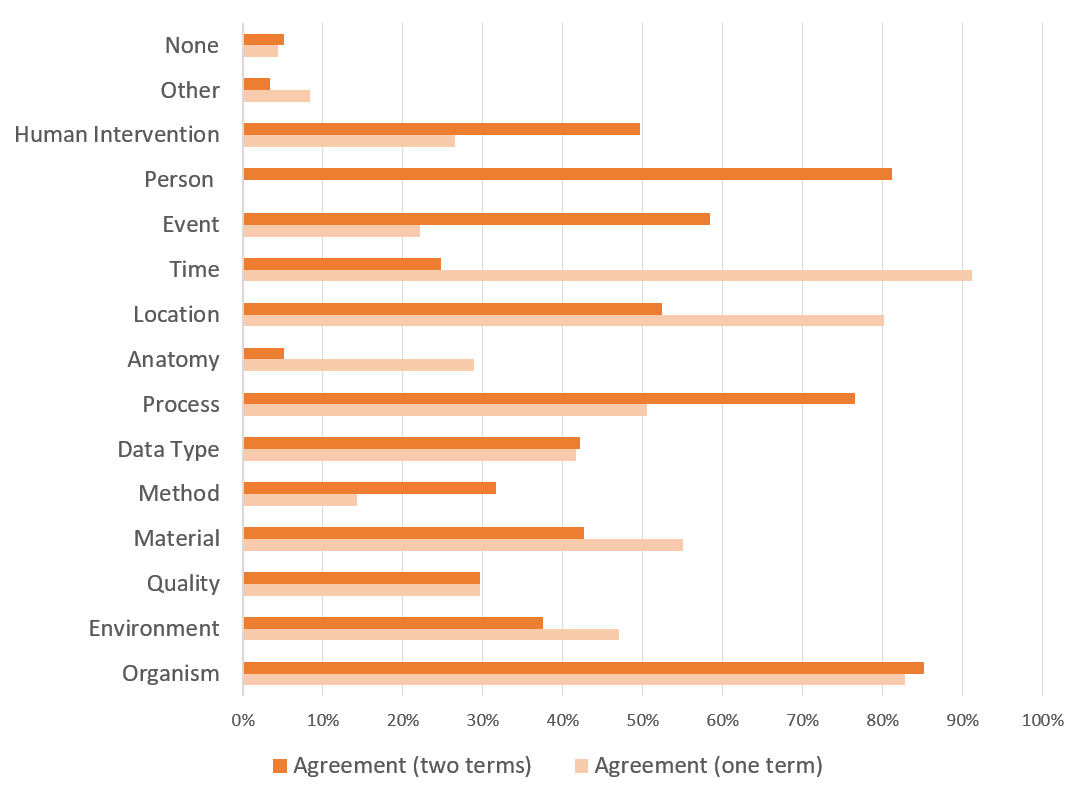}}\hfill
\caption{Fleiss' Kappa values for the individual information categories.} \label{fig:fleisskappa}
\end{adjustwidth}
\end{figure}

With a Fleiss' Kappa of $0.48$
and GWET's AC of $0.51$ the agreement of the annotators over all categories was moderate. Considering the QUALITY correction, the values increase slightly to $0.49$ for Fleiss' Kappa and $0.52$ to GWET's AC. Figure~\ref{fig:fleisskappa}a reveals a more detailed picture. It shows the Fleiss' Kappa for the individual information categories with QUALITY correction. The agreement among the experts was excellent for the categories TIME and ORGANISM and intermediate to good for the categories PERSON \& ORGANIZATION, LOCATION, PROCESS, MATERIALS \& SUBSTANCES and ENVIRONMENT. The experts' agreement for the categories EVENT, HUMAN INTERVENTION, ANATOMY, DATA TYPE, METHOD and QUALITY was fair. This lack of agreement can either point to a different understanding of the categories or might indicate that the categorization of the phrase itself was difficult since some phrases, in particular longer ones with nested entities, were fuzzy and difficult to classify in one category. In the latter case, the annotators were advised not to choose a category for that phrase. Our results show, that for $5\%$ of the phrases at least $2$ annotators did not provide a category. The fraction of phrases where $3$ or more annotators did not choose a category was below $2\%$. This points out that annotators in fact interpreted the categories with poor agreement differently. This correlates with our results regarding the category QUALITY.
For the categories EVENT, HUMAN INTERVENTION, ANATOMY, DATA TYPE, METHOD there is no such evidence. This should be discussed and reconsidered with biodiversity experts.
%

\paragraph{Comparison of short and long artifacts}
We also analyzed the influence of longer artifacts on the result. Table \ref{tableLongShortTerms} presents the $\kappa$ statistic and $GWET's AC$ for artifacts with one term, two terms, three and more terms including the quality correction. As assumed, the longer an artifact is, the more difficult it is to assign an unambiguous category.

\begin{table}[ht!]
\scriptsize
\begin{adjustwidth}{0in}{0in} 
\centering
\caption{
{\bf Annotator's agreement with QUALITY correction overall and for one term, two terms, three terms and more per artifact}}
\begin{tabular}{|p{2.5cm}|p{2.1cm}|p{2.1cm}|p{2.1cm}|p{2.8cm}|}
\hline
& \textbf{Overall} & \textbf{One Term}   & \textbf{Two Terms} & \textbf{$>=$ Three Terms} \\[5pt]\thickhline
$Fleiss' Kappa$  & 0.49 & 0.54 & 0.50  & 0.33 \\[5pt]
\hline
$GWET's AC$ & 0.52 & 0.57 & 0.53 & 0.37 \\[5pt]
\hline
\end{tabular}

\label{tableLongShortTerms}
\end{adjustwidth}
\end{table}

Figure \ref{fig:fleisskappa}b depicts a more detailed picture on the individual categories for artifacts with one and two terms. Since artifacts with three and more terms resulted in coefficients with less than $0.4$, we left them out in this analysis. One-term artifacts got an excellent agreement ($>0.8$) for the categories ORGANISM, TIME and LOCATION and a moderate agreement for ENVIRONMENT, MATERIAL, PROCESS and DATA TYPE. It strikes that PERSON results in a negative value with a poor agreement. Since full person names usually contain two terms, there were no artifacts with one term that could be assigned to PERSON \& ORGANIZATION. However, looking at the results for two terms per artifact, the PERSON category reaches an excellent agreement as well as ORGANISM. Surprisingly, PROCESS ($0.76$) got a substantial agreement for two terms pointing out that biological and chemical processes are obviously mainly defined by two terms. The same effect, a larger agreement for two terms than one term, can also be observed for the categories EVENT and HUMAN INTERVENTION. DATA TYPE got a moderate agreement for one and two terms.

\begin{figure}
\centering
\includegraphics[width=0.99\textwidth]{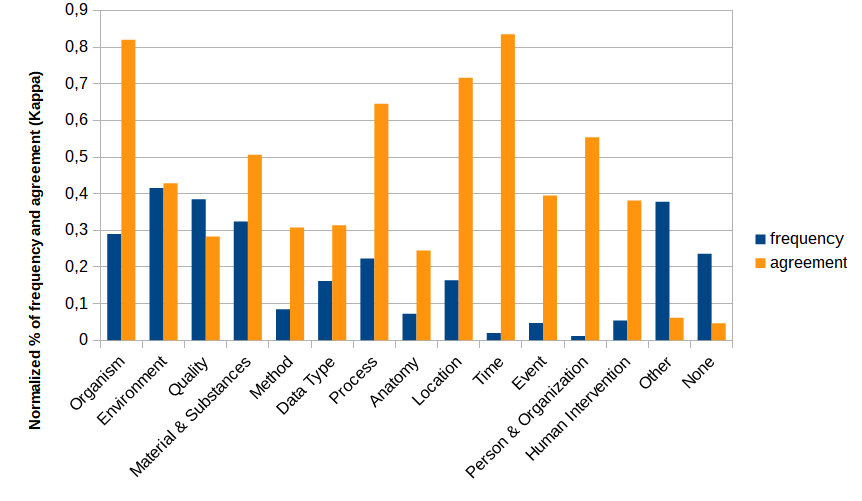}\hfill
\caption{Frequency of category mentions and inter-rater agreement with QUALITY correction.} \label{fig:freqkappa}
\end{figure}

\paragraph{Summary}
All $13$ provided categories were used by the annotators to label the artifacts in the questions. However, what stands out is the high number of the category OTHER in the frequency analysis. For 45\% out of 592 annotations, at least one domain expert did not assign one of the given categories but selected OTHER. That points to missing interests that are not represented by the given classes.
In terms of consensus, seven information categories got a moderate agreement ($>$ 0.4) and five out of these seven were also mentioned very often ($>15\%$), namely ENVIRONMENT (e.g., habitats, climate zone, soil, weather conditions), MATERIAL (e.g., chemicals, geological information), ORGANISM (species, taxonomy), PROCESS (biological and chemical processes) and LOCATION (coordinates, altitude, geographic description) (Figure \ref{fig:freqkappa}). We conclude that these classes are important search interests for biodiversity research.

In comparison to the outcome of the content assessment analysis conducted in the GBIF community \cite{Arino2013} in 2009, the assumption that user interests change over time has been confirmed. Species are still an important category scholars are interested in, however, further important topics for the acquisition and description of ecosystem services are emerging.

\noindent
We are aware that this result is not complete and leaves room for improvement. Some category names were misleading and confused the annotators. That is reflected in fair and bad agreement for some categories such as QUALITY (data parameters measured) or DATA TYPE (nature or genre of the primary data). Here, it should be discussed in the research community how they could be further considered in search, e.g., re-naming or merging of categories. Since the backgrounds of the annotators were quite diverse and no training took place, we  did not expect completeness and perfect agreement. We wanted to get a real, genuine, and unbiased first picture of biodiversity scholars' comprehension when looking for scientific data. In biology and biodiversity research, scholars use a specialized language with diverse content and imprecise and inconsistent naming \cite{Thessen2012,ANANIADOU2004}. Hence, labeling and extracting biological entities remain a challenge. Therefore, our thresholds for agreement ($>$ 0.4) and frequency ($>15\%$) are not as high as in similar studies in bio-medicine.

Concerning the shortened methodology for the evaluation, our assumptions have been confirmed. It saved a lot time that only a few people did the identification of artifacts to be labeled and that domain experts assigned categories, only. On average, domain experts spent between two and three hours for labeling 169 questions. We conclude that our shortened annotation approach is fine for opening up new domains and getting insights in what scholars are interested in. If the aim is to achieve higher agreement per annotation, we recommend training sessions and trial rounds. However, it should be considered that in this case the unbiased feedback gets lost.

For further reuse of the annotated question corpus, our analysis script also produces an XML file with all questions and annotations above a certain agreement threshold that can be set as a parameter. By default, all annotations per question with an agreement above $0.6$ will be returned.

%


\section*{B - Metadata Standards in the Life Sciences}
\label{sec:metadata}

In this section, we describe a selection of  existing metadata standards and investigate whether their  elements  reflect the identified information categories.

\subsection*{Methodology}
Metadata describe scientific primary data such as experiments, tabular data, images, sound and acoustic files in a structured format, e.g., XML or JSON. 
Metadata possess a \emph{schema} stored typically in an XSD file outlining which elements and attributes exist and which of them are mandatory and/or repeatable. Many schemes  employ vocabularies or ontologies to ensure that the metadata use the same names for concepts. In order to become a metadata standard, a schema needs to be formally adopted by a standards' organization such as the International Organization for Standardization, \url{https://www.iso.org}.\\

\noindent
There are a variety of metadata standards for different research fields. Table \ref{tableStand1} presents a list of 13 metadata standards used in data repositories for the Life Sciences. All metadata standards were obtained from \emph{re3data} \cite{re3data}. We filtered for ``Life Sciences'' and retrieved a list of 25 standards. The categories \emph{Other} and \emph{Repository-Developed Metadata Schema} have been left out. The \emph{MIBBI} standard is outdated and has been integrated into \emph{ISA-Tab}, so we left it out, too. All other standards that were used in at least 5 repositories have been selected.

We compared them along the focused \emph{Domain}, \emph{Number of Elements} in the standard and \emph{Mandatory Fields} (Table \ref{tableStand2}). The standards are ranked by the number of data repositories supporting them. The number of elements and required fields were either stated on the standard's website or they were obtained from the schema. We also examined, if there is support for the Semantic Web, namely, if the standard supports RDF or OWL formats. According to the FAIR principles \cite{FAIRPrinciples}, community standards, semantic formats, and ontologies ensure interoperability and data reuse. The last two columns denote whether the standard is still maintained and provide some examples of data repositories that support the respective standard.

\begin{table}
\begin{adjustwidth}{-1.2in}{0in}
\centering
\caption{
{\bf Metadata Schemes in the Life Sciences obtained from \emph{re3data} \cite{re3data}}}

\begin{tabular}{p{9.5cm}p{9.5cm}}
\rowcolor{bananamania}
\textbf{Dublin Core} (\scriptsize \url{http://dublincore.org/documents/dces/})  &
\textbf{DDI} (\scriptsize \url{https://www.ddialliance.org/}) \\
\rowcolor{bananamania}
Dublin Core is a widely used generic metadata standard offering basic fields for the description of research data. &
The DDI (Document, Discover, Interoperate) standard addresses metadata from questionnaires and surveys in the social, behavioral, economic and health sciences.  \\  [5pt]
\rowcolor{bananamania}
\textbf{Data Cite} (\scriptsize \url{https://schema.datacite.org}) &
\textbf{ISO19115} (\scriptsize \url{https://www.iso.org/standard/53798.html}) \\
\rowcolor{bananamania}
Data Cite relates to generic research data and comprises a set of mandatory, recommended and optional properties.  &
The ISO19115 metadata standard includes the identification, the extent, the quality, the spatial and temporal schema, spatial reference, and distribution of digital geographic data. \\ [5pt]
\rowcolor{bananamania}
\textbf{DIF} (\scriptsize \url{https://gcmd.nasa.gov/DocumentBuilder/defaultDif10/guide}) &
\textbf{FDGC/CSDGM} (\scriptsize \url{https://www.fgdc.gov/metadata/csdgm-standard}) \\
\rowcolor{bananamania}
The Directory Interchange Format (DIF) is the US-predecessor of ISO 19115 and focuses on the description of geospatial metadata.
&
The Federal Geographic Data Committee Content Standard for Digital Geospatial Metadata is a legacy national standard for geospatial data developed in the United States. FGDC now encourages its research community to use the international ISO standards. \\ [5pt]
\rowcolor{bananamania}
\textbf{EML} (\scriptsize \url{https://knb.ecoinformatics.org/}) &
\textbf{Darwin Core} (\scriptsize \url{https://dwc.tdwg.org/})\\
\rowcolor{bananamania}
The Ecological Metadata Language (EML) is a series of XML document types that can be used in a modular and extensible manner to document ecological data. &
The Darwin Core standard provides metadata fields for sharing biodiversity data. It is primarily based on taxa, their occurrence in nature and related information. \\ [5pt]
\rowcolor{bananamania}
\textbf{RDF Data Cube} (\scriptsize \url{https://www.w3.org/TR/vocab-data-cube/}) &
\textbf{ISA - Tab} (\scriptsize \url{https://isa-specs.readthedocs.io}) \\
\rowcolor{bananamania}
The RDF Data Cube vocabulary aims to describe statistical data. The model is compatible with the cube model that underlies the Statistical Data and Metadata eXchange standard (SDMX, \url{https://sdmx.org/}), an ISO standard for exchanging and sharing statistical data and metadata among organizations. &
The ISA specification is not a standard but a metadata framework that addresses the description and management of biological experiments. It comprises three core entities to capture experimental metadata: Investigation (the project context), Study (a unit of research) and Assay (analytical measurements).\\ [20pt]
\rowcolor{bananamania}
\textbf{ABCD} (\scriptsize \url{https://github.com/tdwg/abcd})  &
\textbf{CF} (\scriptsize \url{http://cfconventions.org/}) \\
\rowcolor{bananamania}
The ABCD (Access to Biological Collection Data) metadata standards aims to share biological collection data. It offers a variety of metadata fields
to describe specimen and observations, and it is compatible with numerous existing standards.&
The Conventions for Climate and Forecast Metadata (CF) comprise geophysical quantities to describe climate and forecast data. \\ [5pt]
\rowcolor{bananamania}
\textbf{DCAT} (\scriptsize \url{https://www.w3.org/TR/vocab-dcat/}) & \\
\rowcolor{bananamania}
The Data Catalog Vocabulary (DCAT) facilitates interoperability between data catalogs in the web and allows dataset search across sites. &\\ [5pt]
\end{tabular}
\label{tableStand1}
\end{adjustwidth}
\end{table}

\begin{table}
\scriptsize
\begin{adjustwidth}{-1.2in}{0in} 
\centering
\caption{
{\bf Comparison of metadata standards and specifications used in data repositories for Life Sciences. The number in brackets denotes the number of repositories supporting the standard.}}
\begin{tabular}{|M{2.7cm}|M{2.5cm}|M{2cm}|M{2cm}|M{2cm}|M{2cm}|M{3.5cm} |}
\hline
\hline
\textbf{Standard Name} & 
\textbf{Domain} & \textbf{Elements}   & \textbf{Mandatory Elements} & \textbf{Semantic Support} & \textbf{Maintenance} &  \textbf{Examples} \\[10pt]
\thickhline
$Dublin Core (142)$  & 
 general  &	 15	&  No  &  Yes (RDFS) &  Yes 
 &  Pangaea, Dryad, GBIF, Zenodo, Figshare\\[10pt]
\hline
$DDI (74)$  & 
questionnaires and surveys in the social, behavioral, economic, and health sciences  &	 1154 
& 7  &   No &  Yes &  Dataverse\\[10pt]
\hline
$Data Cite (60)$ & 
general research data & 19 (57) & 5 & No & Yes 
& Pangaea, Zenodo, Figshare, Radar\\[10pt]
\hline
$ISO19115 (36)$ & 
geospatial data & N/A & 7 & No  & Yes  & Pangaea, NSF Arctic Data Center, coastMap\\[10pt]
\hline
$FDGC/ CSDGM (34)$ & 
geographic information & 342 & 74 & No & No (1998, last update: 2002)  & Dataverse, NSF Arctic Data Center \\[10pt]
\hline
$EML(21)$ & 
ecological data & N/A & N/A  & No & Yes & GBIF, GFBio, SNSB, Senckenberg, WORMS, NSF Arctic Data Center \\[10pt]
\hline
$DarwinCore (21)$ & 
biodiversity data & 184 
& No & Yes (RDF) & Yes & GFBio, GBIF, VerNET, Atlas of Living Australia, WORMS \\[10pt]
\hline
$RDF Data Cube (18)$ & 
statistical data & 36 & N/A & Yes & Yes  & Dryad (only RDF with DublinCore)
\\[10pt]
\hline
$ISA-Tab (9)$ & 
biological experiments & N/A & Yes (11 blocks)
& Yes & Yes & Data Inra, GigaDB \\[10pt]
\hline
$DIF (7)$ & 
geospatial metadata & 34(219) & 8 
& No & Yes  &  Pangaea, Australian Antarctic Data Center, Marine Environmental Data Section\\[10pt]
\hline
$CF (7)$ & 
climate and forecast & 4798|54|70 (lines in the standard table) & No & No & Yes  & WORMS, NSF Arctic Data Center, coastMap \\[10pt]
\hline
$ABCD (7)$ & 
biological collection data & 1418
& 20 & No &  Yes (ABCD 3.0) &  GBIF, BioCase Network \\[10pt]
\hline
$DCAT (6)$ & 
data catalogs, data sets & 16 & N/A & Yes & Yes & Data.gov.au, European Data Portal \\[10pt]
\hline

\end{tabular}
\begin{flushleft}
\begin{tabular}{l}
N/A denotes that the information was not available
\end{tabular}
\end{flushleft}
\label{tableStand2}
\end{adjustwidth}
\end{table}

\noindent
The standard supported by most repositories is \emph{Dublin Core}, a general metadata standard based on 15 fields, such as contributor, coverage, creator, date, description, format, and identifier. In addition, data repositories utilize further domain-specific standards with richer vocabulary and structure such as \emph{ISO19115} for geospatial data or \emph{EML} for ecological data. The \emph{RDF Data Cube Vocabulary} is not used by any of the data centers.
We suppose, the abbreviation \emph{RDF DC} might lead to a misunderstanding (\emph{DublinCore} instead of \emph{RDF Data Cube}).
All standards provide elements that can be described along the questions: Who? What? Where? When? Why? and How?. In particular, contact person, collection or publication date and location are considered with one or several metadata fields in all standards. In order to describe the main scope of the primary data, all standards offer numerous metadata fields but differ in their granularity. While  simple ones such as \emph{Dublin Core} only offer fields such as title, description, format, and type, standards with more elements such as \emph{EML} or \emph{ABCD} even offer fields for scientific names, methods and data attributes measured. \emph{EML} even allows scholars to define the purpose of the study making it the only standard that supports the \emph{Why} question.
Data reuse and citation also play an important role. As it is demanded by the Joint Declaration of Data Citation Principles \cite{DataCitationPrinciples2014} and practical guidelines for data repositories \cite{fennerData2019}, all standards provide several elements for digital identifiers, license information and citation. In addition, some standards provide elements for data quality checks. For instance, \emph{ISO19115} offers a container for data quality including lineage information and \emph{EML} supports quality checks with the \emph{qualityControl} element. Surprisingly, 52 repositories stated to use own-developed metadata schemes. That indicates that a variety of data repositories is not satisfied with the existing metadata landscape and therefore started developing their own schema. \\

\noindent
For our further analysis, we selected 12 out of the 13 standards shown in Table \ref{tableStand2}. Since \emph{DDI} is a standard that was mainly developed for questionnaires and surveys, we decided not to use it.

\subsection*{Results}

In our second analysis, we compared the information categories with elements of the metadata schemes to figure out, if search interests can be explicitly described with metadata elements.

Our results are presented in Table \ref{tableMetadataResults}. For the sake of completeness, we explored all $13$ categories from the previous analysis but marked the ones with an asterisk that had a fair agreement ($<$ 0.4). The categories are sorted by frequency from left to right. The red color denotes that no element is available in the standard to express the category, orange indicates that only a general field could be used to describe the category and a light-orange cell implies that one or more elements are available in the standard for this search interest.

{\renewcommand{\arraystretch}{1.5}%
\begin{table}[ht!]
\scriptsize
\begin{adjustwidth}{-0in}{0in} 
\caption{
{\bf Comparison of metadata standards and information categories. The categories are sorted by frequency, the asterisk denotes the categories with an agreement less than 0.4}}
\begin{tabular}{|p{2.5cm}|p{0.4cm}|p{0.4cm}|p{0.4cm}|p{0.4cm}|p{0.4cm}|p{0.4cm}|p{0.4cm}|p{0.4cm}|p{0.4cm}|p{0.4cm}|p{0.4cm}|p{0.4cm}|p{0.4cm}|}
\hline
& \rotatebox[origin=c]{90}{\textbf{Environment}}   & \rotatebox[origin=c]{90}{\textbf{Quality*}} & \rotatebox[origin=c]{90}{\textbf{Material}} & \rotatebox[origin=c]{90}{\textbf {Organism}} & \rotatebox[origin=c]{90}{\textbf{Process}} &  \rotatebox[origin=c]{90}{\textbf{Location}} &  \rotatebox[origin=c]{90}{\textbf{Data Type*}} &  \rotatebox[origin=c]{90}{\textbf{Method*}} & \rotatebox[origin=c]{90}{\textbf{Anatomy*}} &  \rotatebox[origin=c]{90}{\textbf{Human Intervention*}} &  \rotatebox[origin=c]{90}{\textbf{Event*}} &  \rotatebox[origin=c]{90}{\textbf{Time}} & \rotatebox[origin=c]{90}{\textbf{Person}}\\ \thickhline
$Dublin Core$  & \cellcolor{YellowOrange} & \cellcolor{YellowOrange} & \cellcolor{YellowOrange} & \cellcolor{YellowOrange} & \cellcolor{YellowOrange} & \cellcolor{Yellow} & \cellcolor{Yellow} & \cellcolor{YellowOrange}& \cellcolor{YellowOrange} & \cellcolor{YellowOrange} & \cellcolor{YellowOrange} & \cellcolor{Yellow} & \cellcolor{Yellow}\\
\hline
$Data Cite$ & \cellcolor{YellowOrange} & \cellcolor{YellowOrange} & \cellcolor{YellowOrange} & \cellcolor{YellowOrange} & \cellcolor{YellowOrange} & \cellcolor{Yellow} & \cellcolor{Yellow} & \cellcolor{YellowOrange} & \cellcolor{YellowOrange} & \cellcolor{YellowOrange} & \cellcolor{YellowOrange} & \cellcolor{Yellow} & \cellcolor{Yellow}\\
\hline
$ISO19115$  & \cellcolor{YellowOrange} & \cellcolor{YellowOrange} & \cellcolor{YellowOrange} & \cellcolor{YellowOrange} &  \cellcolor{YellowOrange} & \cellcolor{Yellow} & \cellcolor{Yellow} & \cellcolor{YellowOrange} & \cellcolor{YellowOrange} & \cellcolor{YellowOrange} &  \cellcolor{YellowOrange} & \cellcolor{Yellow} & \cellcolor{Yellow}\\ 
\hline
$FDGC/ CSDGM$ & \cellcolor{YellowOrange} & \cellcolor{Yellow} & \cellcolor{YellowOrange} & \cellcolor{YellowOrange} &\cellcolor{YellowOrange} & \cellcolor{Yellow} & \cellcolor{Yellow} & \cellcolor{Yellow} & \cellcolor{YellowOrange} & \cellcolor{YellowOrange} & \cellcolor{YellowOrange} & \cellcolor{Yellow} & \cellcolor{Yellow}\\ 
\hline
$EML$  &  \cellcolor{Yellow} & \cellcolor{Yellow}& \cellcolor{YellowOrange} & \cellcolor{Yellow} & \cellcolor{YellowOrange} & \cellcolor{Yellow} & \cellcolor{Yellow} & \cellcolor{Yellow} & \cellcolor{YellowOrange} & \cellcolor{YellowOrange} & \cellcolor{YellowOrange} & \cellcolor{Yellow} & \cellcolor{Yellow}\\
\hline
$DarwinCore$ & \cellcolor{Yellow} & \cellcolor{Yellow} & \cellcolor{BrickRed} & \cellcolor{Yellow} & \cellcolor{BrickRed} & \cellcolor{Yellow} & \cellcolor{Yellow} & \cellcolor{Yellow} & \cellcolor{BrickRed} & \cellcolor{BrickRed} & \cellcolor{Yellow} & \cellcolor{Yellow} & \cellcolor{Yellow}\\
\hline
$RDF Data Cube$ & \cellcolor{YellowOrange} & \cellcolor{Yellow} & \cellcolor{YellowOrange} & \cellcolor{YellowOrange} &\cellcolor{YellowOrange} & \cellcolor{YellowOrange} & \cellcolor{YellowOrange} & \cellcolor{YellowOrange} & \cellcolor{YellowOrange} & \cellcolor{YellowOrange} & \cellcolor{YellowOrange} & \cellcolor{Yellow} & \cellcolor{Yellow}\\
\hline
$ISA-Tab$ & \cellcolor{Yellow} & \cellcolor{Yellow} & \cellcolor{Yellow} & \cellcolor{Yellow} & \cellcolor{Yellow} & \cellcolor{Yellow} & \cellcolor{Yellow} &\cellcolor{Yellow} & \cellcolor{Yellow} & \cellcolor{BrickRed} & \cellcolor{BrickRed}& \cellcolor{Yellow} & \cellcolor{Yellow}\\
\hline
$DIF$ & \cellcolor{YellowOrange} & \cellcolor{Yellow} & \cellcolor{YellowOrange} & \cellcolor{YellowOrange} & \cellcolor{YellowOrange} & \cellcolor{Yellow} & \cellcolor{Yellow} & \cellcolor{Yellow} & \cellcolor{YellowOrange} & \cellcolor{YellowOrange} & \cellcolor{YellowOrange} & \cellcolor{Yellow} & \cellcolor{Yellow}\\
\hline
$CF$ & \cellcolor{Yellow} & \cellcolor{Yellow} & \cellcolor{BrickRed} & \cellcolor{BrickRed} & \cellcolor{BrickRed} & \cellcolor{Yellow} & \cellcolor{BrickRed} & \cellcolor{BrickRed} & \cellcolor{BrickRed} & \cellcolor{BrickRed} & \cellcolor{BrickRed} & \cellcolor{BrickRed} & \cellcolor{BrickRed}\\
\hline
$ABCD$ & \cellcolor{YellowOrange} & \cellcolor{Yellow} & \cellcolor{BrickRed} & \cellcolor{Yellow} & \cellcolor{BrickRed} & \cellcolor{Yellow} & \cellcolor{Yellow} & \cellcolor{Yellow} & \cellcolor{Yellow} & \cellcolor{BrickRed} & \cellcolor{BrickRed} & \cellcolor{Yellow} & \cellcolor{Yellow}\\
\hline
$DCAT$ & \cellcolor{YellowOrange} & \cellcolor{YellowOrange} & \cellcolor{YellowOrange} & \cellcolor{YellowOrange} & \cellcolor{YellowOrange} & \cellcolor{YellowOrange} & \cellcolor{Yellow} & \cellcolor{YellowOrange} & \cellcolor{YellowOrange} & \cellcolor{YellowOrange} & \cellcolor{YellowOrange} & \cellcolor{YellowOrange} & \cellcolor{Yellow}\\
\hline

\end{tabular}
\begin{flushleft}
\textbf{Table key}
\\
\begin{tabular}{llllll}
\textcolor{BrickRed}{$\blacksquare$} & Not provided & \textcolor{YellowOrange}{$\blacksquare$} & Unspecific (generic element) &
\textcolor{Yellow}{$\blacksquare$} & Available (one or more elements)
\end{tabular}
\end{flushleft}
\label{tableMetadataResults}
\end{adjustwidth}
\end{table}
}

\noindent
There is no schema that covers all categories. Since the interests are obtained from scholars with various and heterogeneous research backgrounds, this was also not to be expected. Some standards such as \emph{ABCD} or \emph{DarwinCore} are discipline-specific and therefore, mainly provide elements that support the respective domain (e.g., collection data).

Apart from HUMAN INTERVENTION, all categories are covered by different metadata schemes. In particular, \emph{ISA-Tab} followed by \emph{ABCD}, \emph{DarwinCore} and \emph{EML} are frameworks and metadata schemes with elements that cover most of the search interests of biodiversity researchers. \emph{EML} provides numerous fields to describe ecological data including elements for environmental information (\texttt{studyAreaDescription}), species (\texttt{taxonomicCoverage}) and research methods used (\texttt{methods}). However, important search preferences such as materials (including chemicals) and biological and chemical processes are only explicitly supported by \emph{ISA-Tab}.
Widely used general standards such as \emph{DublinCore} or \emph{DataCite} offer at least a general field (\texttt{dc:subject}, \texttt{subject}) that could be used to describe the identified search categories. In \emph{DublinCore}, at least one metadata field each is provided to describe geographic information, e.g., where the data have been collected (LOCATION), the type of the data (DATA TYPE), the creator and contributor (PERSON \& ORGANIZATION) and when it was collected or published (TIME). However, one field is often not enough to distinguish if the provided field is for instance a collection date or publication date, or if the creator of the dataset is also the same person that collected the data. In contrast, \emph{DataCite} provides individual fields for publication year and the \texttt{date} field can be used with \texttt{dateType="Collected"} to specify a collection date. The metadata field \texttt{contributor} can also be extended with a type to indicate whether the contact details belong to the data collector or the project leader. Bounding box elements are also provided to enter geographic coordinates (LOCATION).

The question that still remains to be answered is whether these detailed metadata standards are actually used by data repositories.
\section*{C - Metadata Usage in Selected Data Repositories}
\label{sec:dataRepo}

In the following analysis, we examine what metadata standards are used in selected data repositories and how many schema elements are actually filled. In a second part, we explore, if descriptive fields of selected files contain data that might be relevant for information seekers.


\subsection*{Methodology}
Scholarly publishers increasingly demand  scientific data to be submitted along with  publications. Since publishers usually do not host the data on their own, they ask scholars to upload the data at one of the repositories for their research domain. According to \emph{Nature's} list of recommended data repositories \cite{NatureRepoList}, we selected five archives for our further analysis: three generalist ones (\emph{Dryad}, \emph{Figshare} and \emph{Zenodo}) and two domain-specific ones (\emph{PANGAEA} - environmental data, \emph{GBIF} - taxonomic data). In the biodiversity projects we are involved in, scholars also often mention these repositories as the ones they mainly use.

\paragraph{OAI-PMH Harvesting}
The Open Archives Initiative Protocol for Metadata Harvesting (OAI-PMH) is a client/server architecture primarily developed for providing and consuming metadata. Data repositories are required to expose their metadata in $Dublin Core$ metadata format and may also support other metadata formats. Metadata consumers, e.g., other institutions or data portals, harvest that data via the provided services on the OAI-PMH server in order to integrate  or reuse it in their services. The OAI-PMH protocol comprises a set of six services that are accessible via HTTP. Requests for metadata can be based on a date stamp range or can be restricted to named sets defined by the provider. \\

\noindent
We parsed all available metadata from \emph{Figshare}, \emph{Dryad}, \emph{GBIF}, \emph{PANGAEA} and \emph{Zenodo} in May 2019 via their respective \emph{OAI-PMH} interfaces. \emph{GBIF} only offers the metadata of their datasets in the \emph{OAI-PMH} interface. The individual occurrence records, which are provided in \emph{Darwin Core} metadata schema \cite{Gaji2013} and belong to a dataset, are  available in the search, only. Hence, we only analyzed the metadata of the datasets.

Our script parses the metadata fields of all public records per metadata schema for each of the selected data repositories (Table \ref{metadata_formats}). Apart from the metadata standards introduced in the previous section, a few more standards appear in this list. \emph{OAI-DC} is an abbreviation for $DublinCore$, a mandatory standard in OAI-PMH interfaces. \emph{QCD} means \emph{qualified DublinCore} and denotes an extended $DublinCore$  extending or refining the $15$ core elements. \emph{ORE} (The Open Archives Initiative Object Reuse and Exchange (OAI-ORE)) is a standard for exchanging aggregations of web resources. It can be used together with other semantic standards such as RDF to group individual web resources. We also considered \emph{Pan-MD}, a metadata schema developed by \emph{PANGAEA}. It extends $DublinCore$ with more fine-grained geographic information such as bounding boxes or adds information on data collection. The latter can range from projects, parameters, methods, and sensors to taxonomy or habitats.

\begin{table}
\small
	\begin{adjustwidth}{0in}{0in}
			\caption{\bf{Metadata schemes offered by selected data repositories in their OAI-PMH interfaces.}}
			\begin{tabular}{|c | c | c | c | c|}
			\hline
				\textbf{Dryad} & \textbf{GBIF} & \textbf{PANGAEA} & \textbf{Zenodo} & \textbf{Figshare} \\
				\hline \hline
				\href{http://api.datadryad.org/oai/request?verb=ListRecords&metadataPrefix=mets}{METS}&
				\href{http://api.gbif.org/v1/oai-pmh/registry?verb=ListRecords&metadataPrefix=eml}{EML} &
				\href{http://ws.pangaea.de/oai/provider?verb=ListRecords&metadataPrefix=datacite3}{DATACITE3} &
				\href{https://zenodo.org/oai2d?verb=ListRecords&metadataPrefix=datacite}{DATACITE} &
				\href{https://api.figshare.com/v2/oai?verb=ListRecords&metadataPrefix=cerif}{CERIF} \\
				\rule{0ex}{3ex} \href{http://api.datadryad.org/oai/request?verb=ListRecords&metadataPrefix=oai\_dc}{OAI-DC} &
				\href{http://api.gbif.org/v1/oai-pmh/registry?verb=ListRecords&metadataPrefix=oai\_dc}{OAI-DC} &
				\href{http://ws.pangaea.de/oai/provider?verb=ListRecords&metadataPrefix=dif}{DIF} &
				\href{https://zenodo.org/oai2d?verb=ListRecords&metadataPrefix=datacite3}{DATACITE3} &
				\href{https://api.figshare.com/v2/oai?verb=ListRecords&metadataPrefix=mets}{METS} \\
				\rule{0ex}{3ex} \href{http://api.datadryad.org/oai/request?verb=ListRecords&metadataPrefix=ore}{ORE}
				&
				&
				\href{http://ws.pangaea.de/oai/provider?verb=ListRecords&metadataPrefix=iso19139}{ISO19139} &
				\href{https://zenodo.org/oai2d?verb=ListRecords&metadataPrefix=datacite4}{DATACITE4} &
				\href{https://api.figshare.com/v2/oai?verb=ListRecords&metadataPrefix=oai\_datacite}{OAI-DATACITE} \\
				\rule{0ex}{3ex} \href{http://api.datadryad.org/oai/request?verb=ListRecords&metadataPrefix=rdf}{RDF} &
				&
				\href{http://ws.pangaea.de/oai/provider?verb=ListRecords&metadataPrefix=iso19139.iodp}{ISO19139.IODP} &
				\href{https://zenodo.org/oai2d?verb=ListRecords&metadataPrefix=marcxml}{MARCXML} &
				\href{https://api.figshare.com/v2/oai?verb=ListRecords&metadataPrefix=oai\_dc}{OAI-DC} \\
				\rule{0ex}{3ex}
				&
				&
				\href{http://ws.pangaea.de/oai/provider?verb=ListRecords&metadataPrefix=oai\_dc}{OAI-DC} &
				\href{https://zenodo.org/oai2d?verb=ListRecords&metadataPrefix=marc21}{MARC21} &
				\href{https://api.figshare.com/v2/oai?verb=ListRecords&metadataPrefix=qdc}{QDC} \\
				\rule{0ex}{3ex}
				&
				&
				\href{http://ws.pangaea.de/oai/provider?verb=ListRecords&metadataPrefix=pan\_md}{PAN-MD} &
				\href{https://zenodo.org/oai2d?verb=ListRecords&metadataPrefix=oai\_datacite}{OAI-DATACITE} &
				\href{https://api.figshare.com/v2/oai?verb=ListRecords&metadataPrefix=rdf}{RDF} \\
				\rule{0ex}{3ex}
				 &
				 &
				 &
				 \href{https://zenodo.org/oai2d?verb=ListRecords&metadataPrefix=oai\_datacite3}{OAI-DATACITE3} & \\
				\rule{0ex}{3ex}
				&
				&
				&
				\href{https://zenodo.org/oai2d?verb=ListRecords&metadataPrefix=oai\_dc}{OAI-DC} & \\
				\hline \hline
			\end{tabular}
		
			\label{metadata_formats}
	\end{adjustwidth}
\end{table}

\emph{MARC21}, \emph{MARCXML} and \emph{METS} are metadata standards that are mainly used for bibliographic data in digital libraries. Hence, we left them out of our further explorations. We also did not consider the \emph{Common European Research Information Format (CERIF)} and \emph{ORE} as they are not focused on describing primary data but research entities and their relationships and grouping web resources, respectively. However, we decided to permit all available repository-developed schemes for Life Sciences such as \emph{Pan-MD} in order to get an impression how repositories extend metadata descriptions. \\

%

\noindent
Per metadata file we inspected which elements of the metadata standards are used, and we saved their presence (1) or non-presence (0). The result is a csv file per metadata schema that contains dataset IDs and metadata elements used. All generated files are stored in separate folders per repository and metadata format. Each request to a repository returns an XML body that includes several metadata files as records. Each record is separated in two sections, a header and a metadata section. The header section comprises general information such as example ID of the record and a date stamp. The metadata section contains elements of the metadata schema, e.g., the name of the contributors, abstract and publication year. Unused metadata fields are not included in the response. We saved a boolean value encoding whether a metadata field was used or not. The source code and a documentation on how to use it is available in our $GitHub$ repository. \\

\begin{table}
\scriptsize
	\begin{adjustwidth}{-1.3in}{0in}
	\caption{\bf{The date stamps used for each metadata standard and their descriptions obtained from the standard's website.}}
	\begin{center}
	\resizebox{1.5\textwidth}{!}{\begin{tabular}{|p{4.5cm}|p{2.25cm}|p{4.5cm}|p{5.5cm}|}
	  \hline
		\textbf{Format} & \textbf{Element} & \textbf{URL} & \textbf{Description} \\
		\hline \hline
		\emph{OAI-DC/METS/QDC\newline RDF(Dryad)} & dc:date & \url{http://www.dublincore.org/specifications/dublin-core/dces/} & ``A point or period of time associated with an event in the lifecycle of the resource.'' \\ \hline
		\emph{EML} & pubDate & \url{https://knb.ecoinformatics.org/external//emlparser/docs/eml-2.1.1/eml-resource.html\#pubDate} & ``The 'pubDate' field represents the date that the resource was published.'' \\ \hline
		\emph{DATACITE3/4\newline OAI-DATACITE/3} & publicationYear & \url{https://support.datacite.org/docs/schema-40} & ``The year when the data was or will be made publicly available.'' \\ \hline
		\emph{DIF} & DIF\_Creation-\newline\_Date & \url{https://gcmd.gsfc.nasa.gov/DocumentBuilder/defaultDif10/guide/metadata\_dates.html} & `` refers to the date the data was created'' \\ \hline
		\emph{ISO19139/ISO19139.iodp} & gco:DateTime & \url{https://geo-ide.noaa.gov/wiki/index.php?title=ISO\_Dates} & (CI-DataTypeCode=publication), publication Date \\ \hline
		\emph{PAN-MD} & md:dateTime & \url{http://ws.pangaea.de/schemas/pangaea/MetaData.xsd} & publication date (contact to data repository) \\ \hline
		\emph{RDF(Figshare)} & vivo:datePublished & \url{https://codemeta.github.io/terms/} & ''Date of first broadcast/publication''\newline \\ \hline
	\end{tabular}}
	\label{datestamps}
	\end{center}
	\end{adjustwidth}
\end{table}

\noindent
For our further consideration, we wanted to obtain a publication date of each downloaded dataset to inspect how many datasets have been published over the years in which format per data repository. Unfortunately, a publication date is not provided in all metadata schemes. Therefore, we looked up each date related field in the schema and used the one that is (based on the description) the closest to a publication date. Table \ref{datestamps} depicts all date stamps utilized and their descriptions. If the respective date stamp was not found in a dataset or was empty, we left the dataset out in the following analysis.

\paragraph{Content Analysis:}
General, descriptive metadata fields such as `title', `description' or `abstract', and `subject' might contain relevant data that are interesting for information seekers. Using conventional retrieval techniques, this data is only accessible in a full text search and if the entered query terms exactly match a term in the dataset.
Hence, we aim to explore what information is available in general, descriptive metadata fields.

In a first step, we downloaded descriptive metadata fields, namely, \texttt{dc:title}, \texttt{dc:description} and \texttt{dc:subject} in \emph{OAI-DC} format from all repositories in October and November 2019. Parallel to the download, we collected the keywords used in the subject field and counted their presence in a separate csv file.

In order to further inspect the content with Natural Language Processing (NLP) tools, we selected a subset of representative datasets. We limited the amount to 10,000 datasets per repository as the processing of textual resources is time-consuming and resource-intensive.
A variety of applications have been developed to determine Named Entities (NE) such as geographic locations, persons and dates. Thessen et. al \cite{Thessen2012} explored the suitability of existing NLP applications for biodiversity research. Their outcome reveals that current text mining systems, which were mainly developed for the biomedical domain, are able to discover biological entities such as species, genes, proteins and enzymes. Further relevant entity types such as habitats, data parameters or processes are currently not supported by existing taggers. Thus, we  concentrated on the extraction of entity types that (a) correspond to the identified search interests and for which (b) text mining pipelines are available. We used the text mining framework GATE \cite{GATE} and its ANNIE pipeline \cite{ANNIE} as well as the OrganismTagger \cite{Naderi2011} to extract geographic locations, persons, organizations and organisms.

\subsection*{Results}
\label{sec:resQ}

The overall statistics are presented in Table \ref{DataRepoStatistics}. At first, we inspected the fields concerning a publication date in a valid format. We could not use all harvested datasets as for some metadata files publication dates were not available. \emph{Dryad} had a large number of datasets with a status ``Item is not available'', which we left out, too. The number in brackets denotes the amount of datasets we used for the following considerations.
What stands out is that most repositories provide general standards, only \emph{Pangaea} and \emph{GBIF} utilize discipline-specific metadata schemes.
$Dryad$ and $Figshare$ already provide metadata in semantic formats such as RDF. In addition, $Figshare$ offers \emph{Qualified Dublin Core (QDC)}, an extended \emph{Dublin Core} that allows the description of relations to other data sources.

\begin{figure}
\centering
\includegraphics[width=0.5\textwidth]{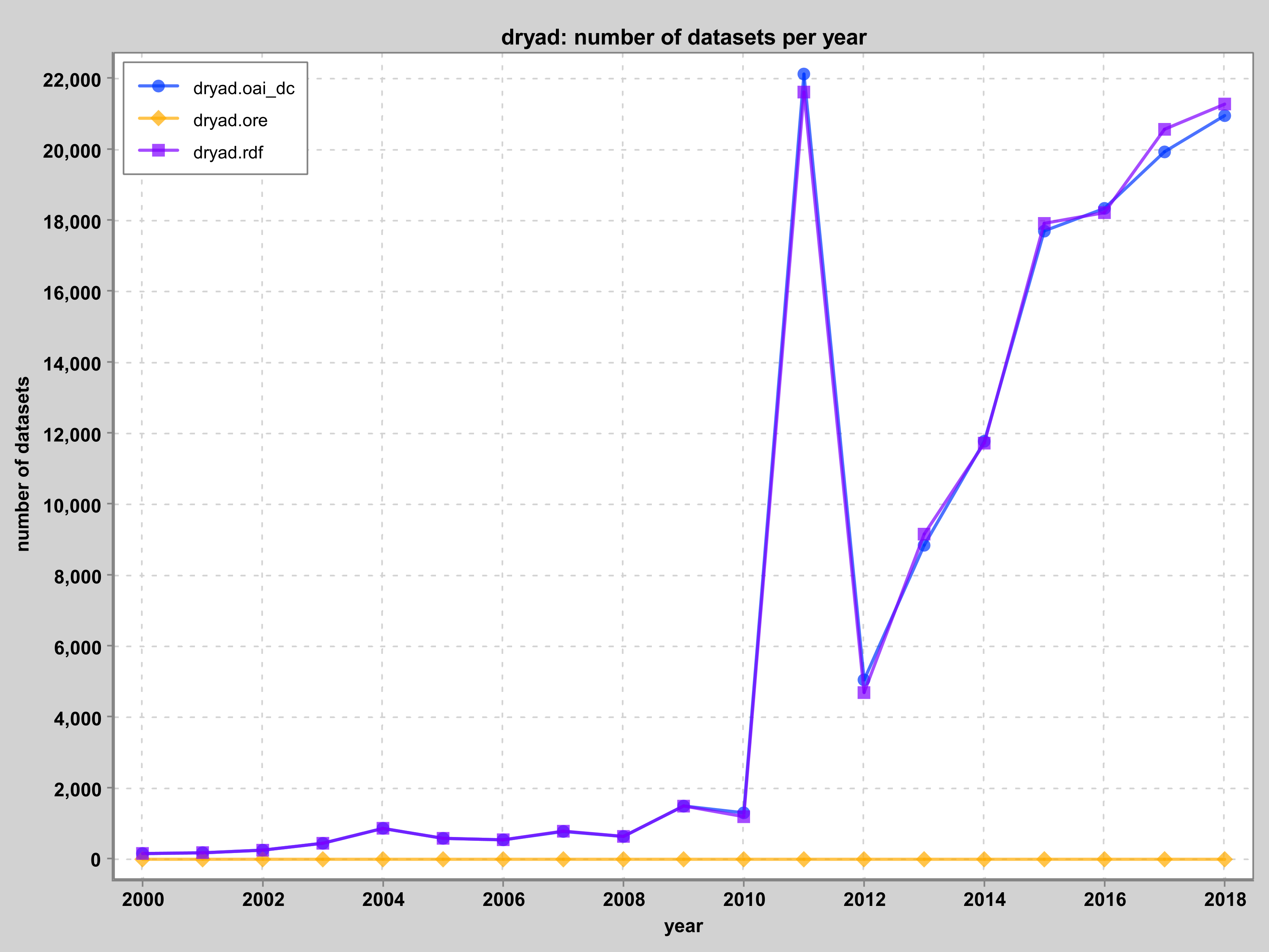}\hfill
\includegraphics[width=0.5\textwidth]{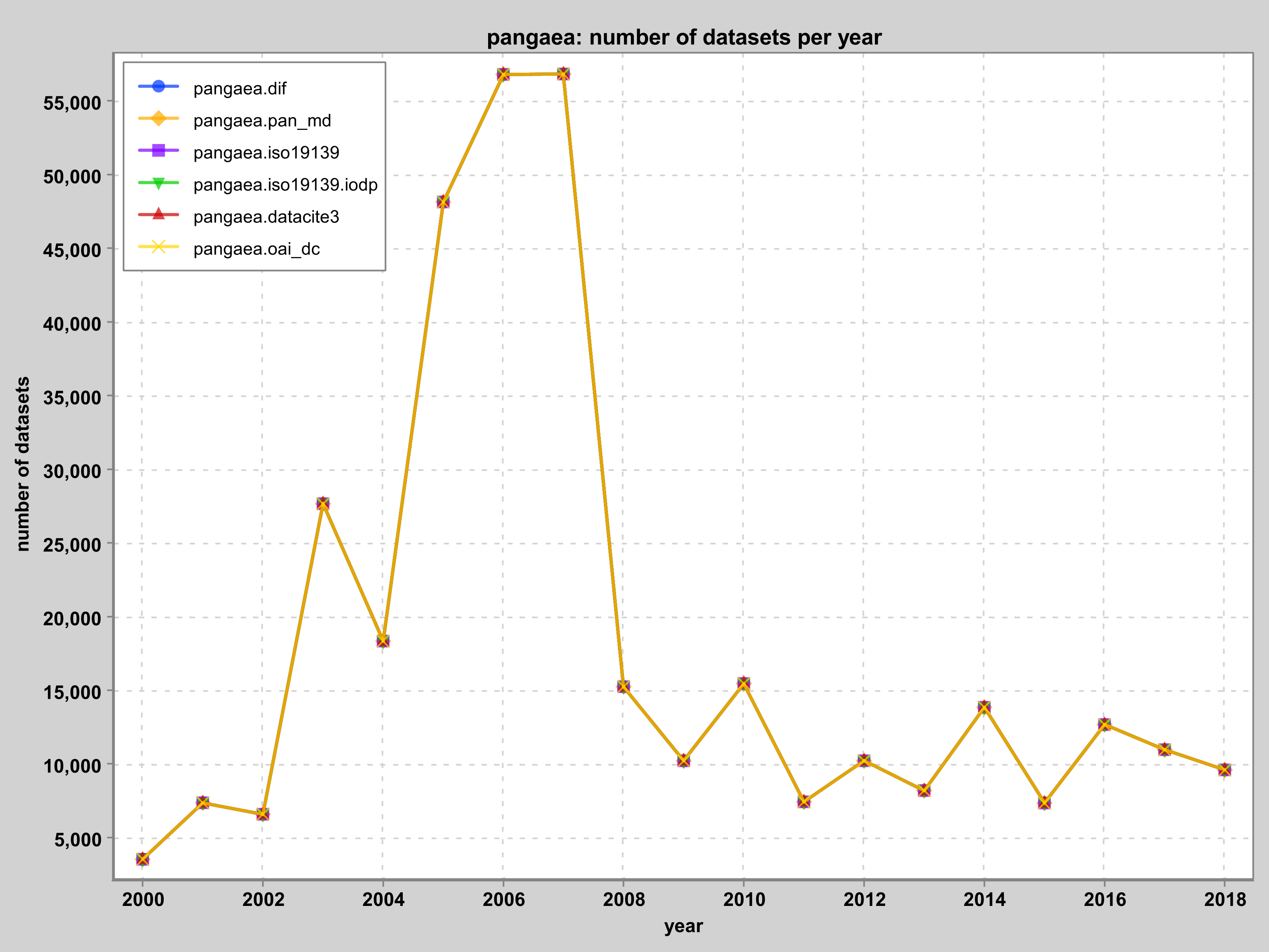}\hfill
\includegraphics[width=0.5\textwidth]{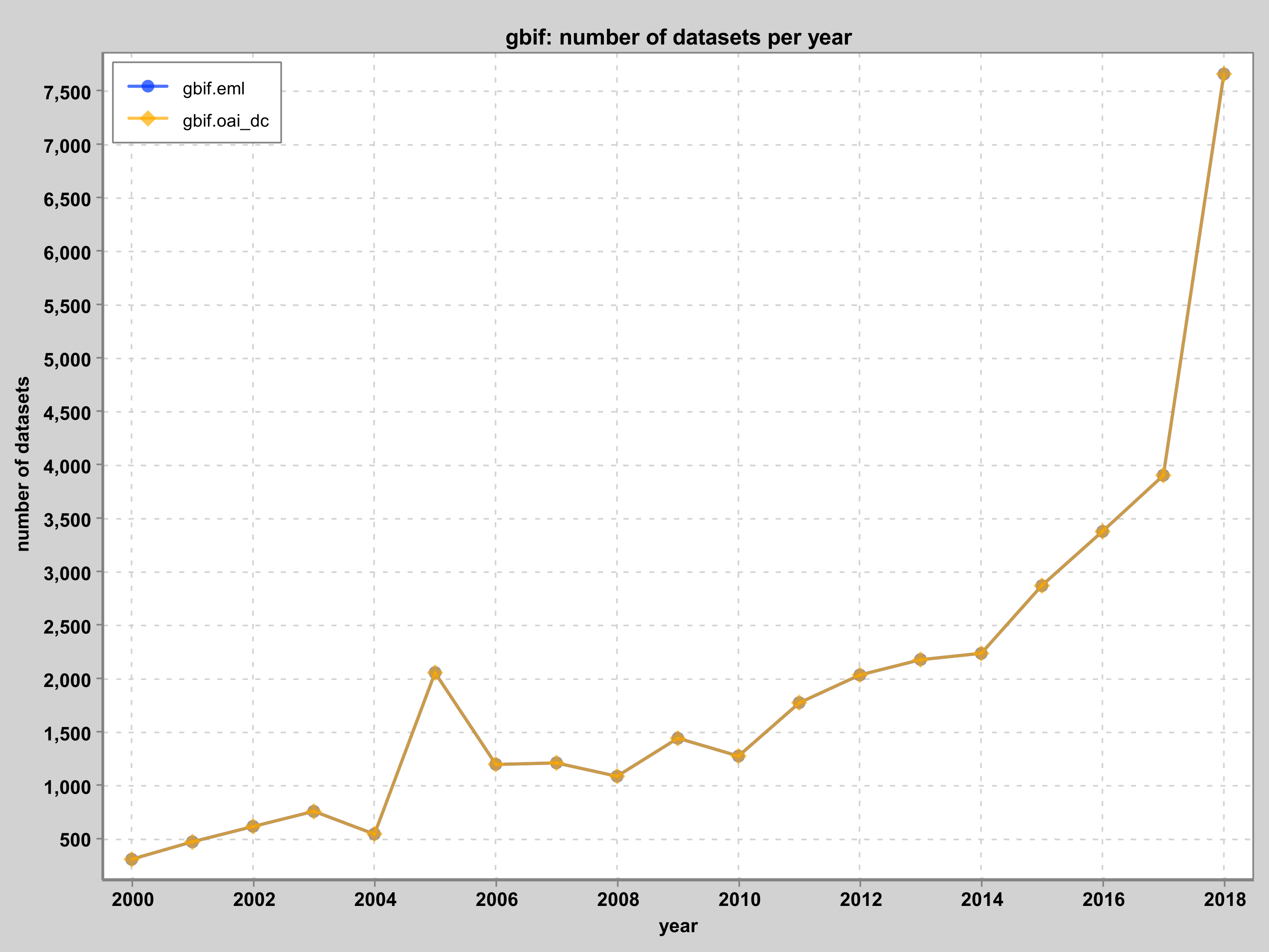}\hfill
\includegraphics[width=0.5\textwidth]{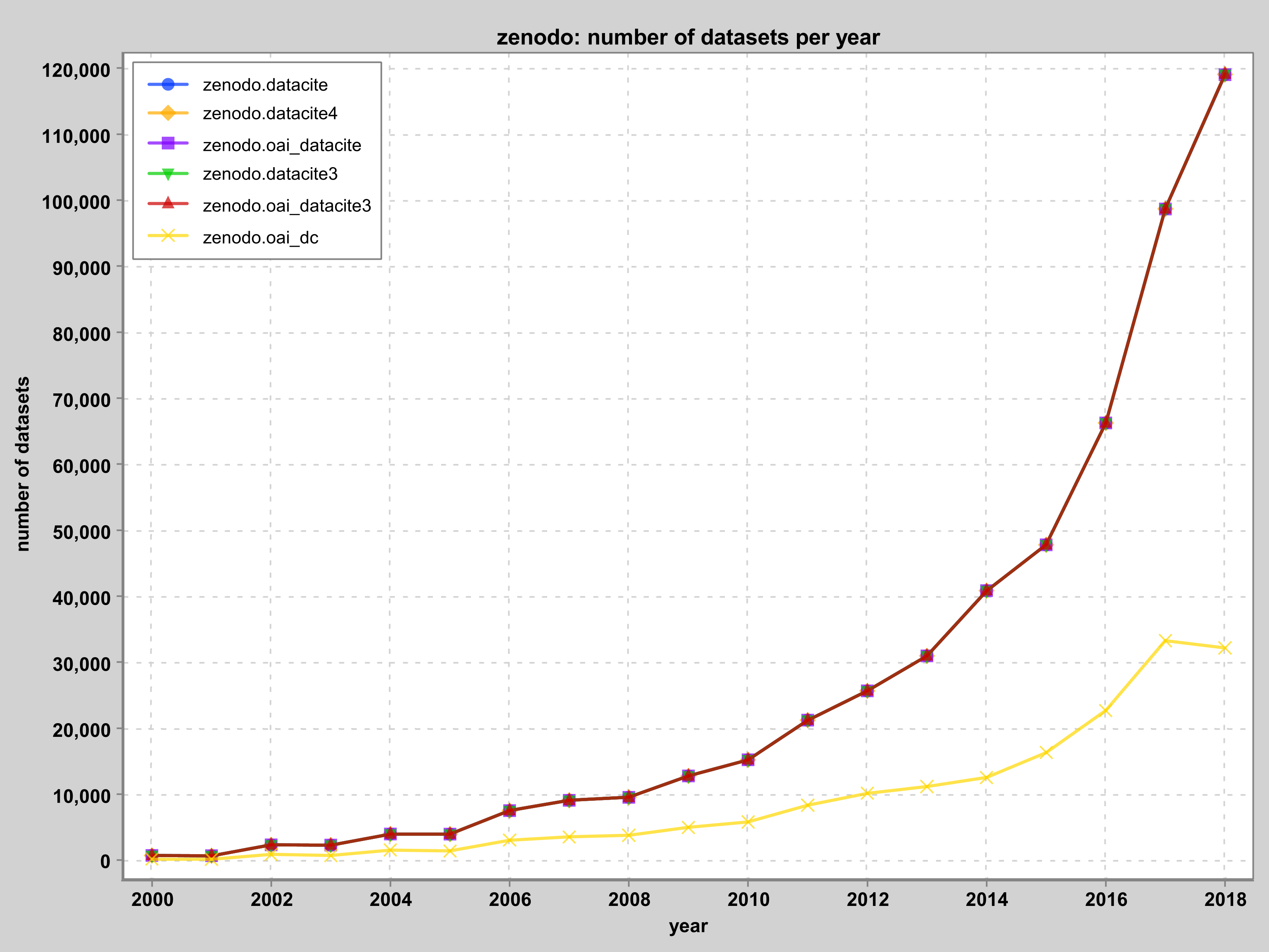}\hfill
\includegraphics[width=0.5\textwidth]{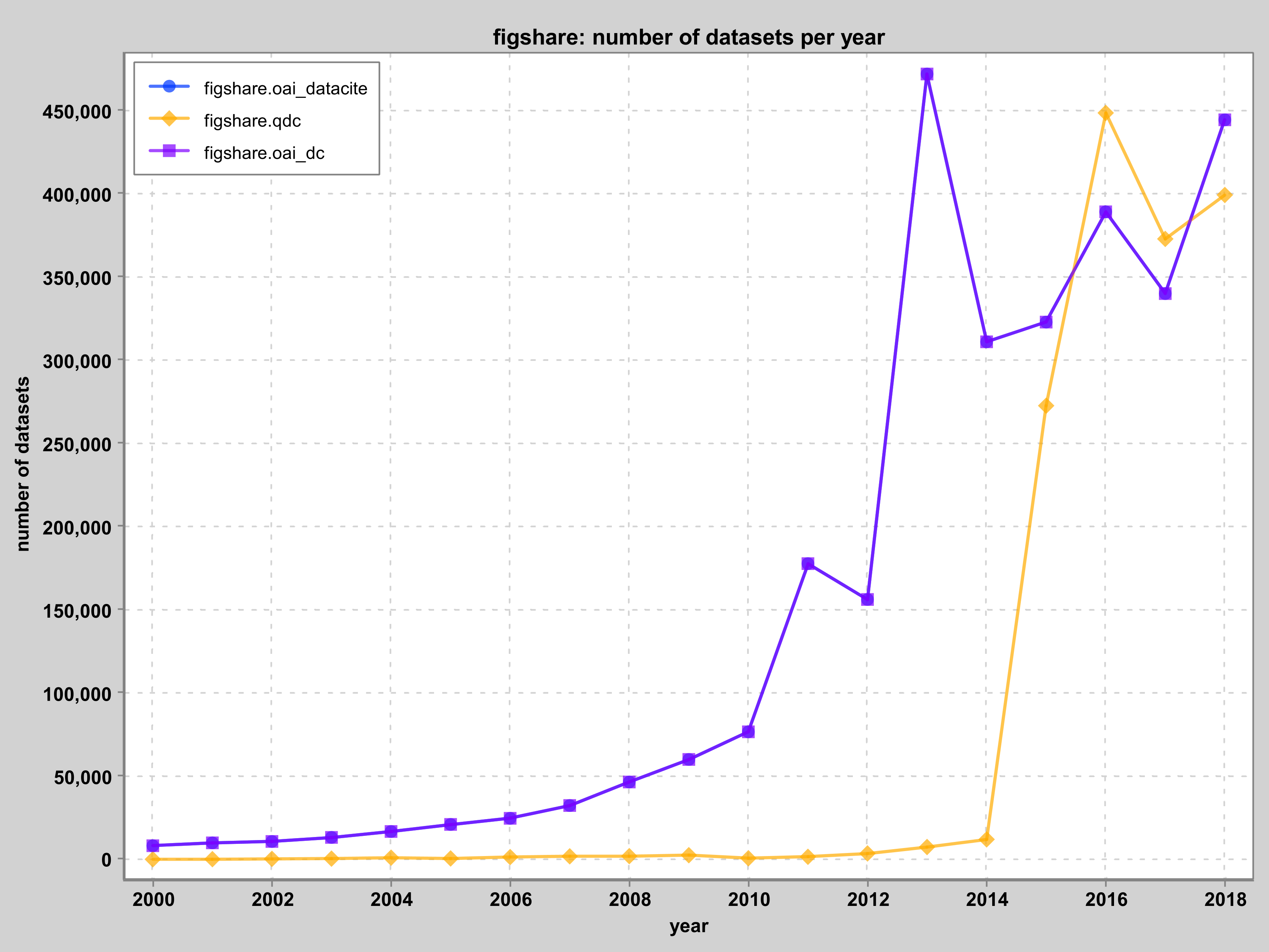}\hfill
\includegraphics[width=0.5\textwidth]{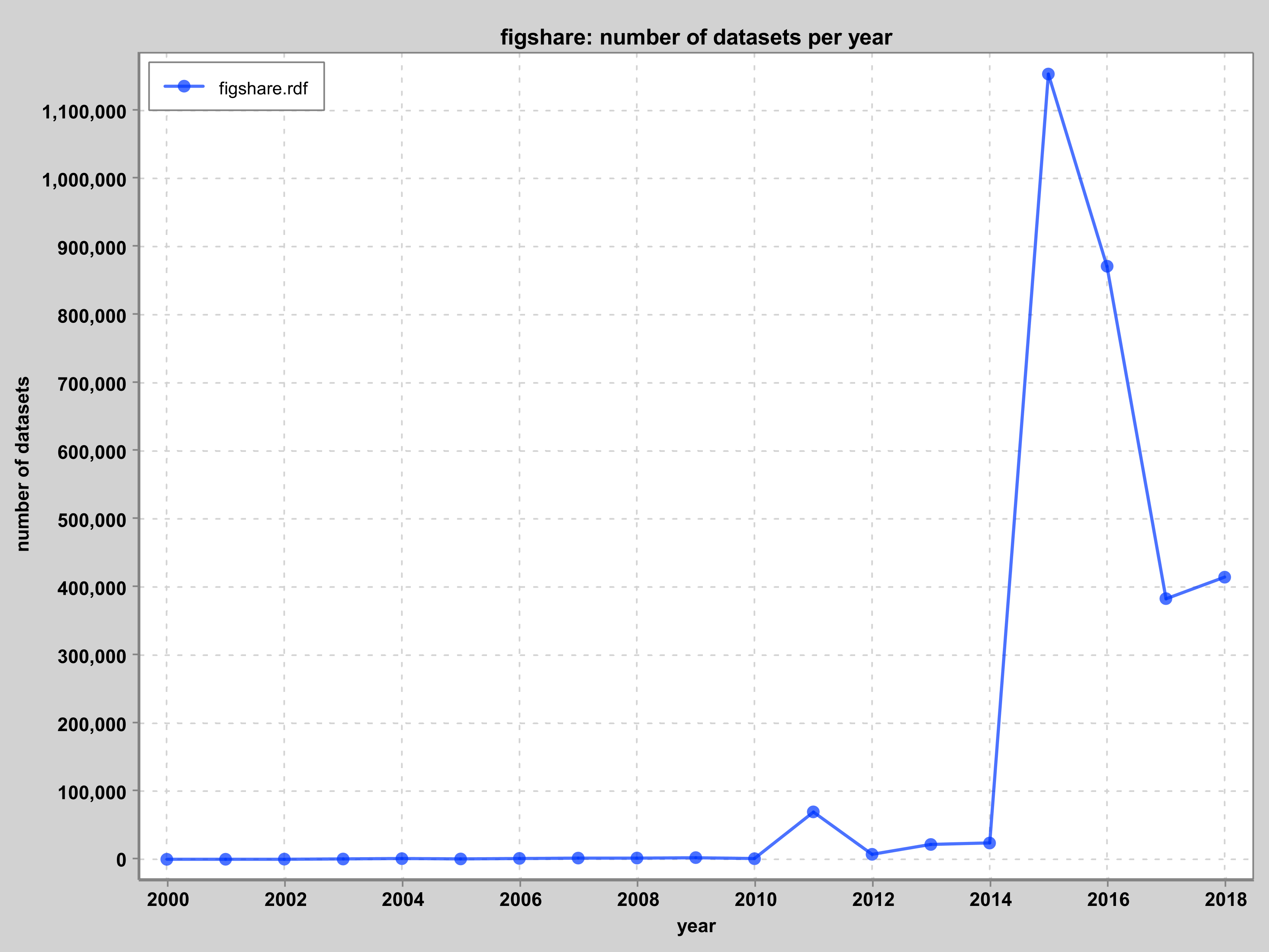}\hfill
\caption{Timelines for all repositories presenting the number of datasets per metadata schema offered. \emph{Figshare's} timeline for RDF was computed separately as the data are too large to process it together with the other metadata formats.} \label{fig:timelines}
\end{figure}

\paragraph{Timelines}
Based on the given publication dates, we computed timelines (Figure \ref{fig:timelines}) for the introduction of the various standards over time per data repository. The code and all charts are available in the repository. As \emph{Dryad} provides several \texttt{dc:date} elements in the metadata, we used the first available date entry as publication date for the timeline chart.

Per repository, the timelines for the different metadata formats are almost identical. Obviously, when introducing a new metadata format, publication dates were adopted from existing metadata formats. Only \emph{Figshare} uses new date stamps when a new metadata format is provided. For instance, \emph{Figshare}'s timeline shows that QDC and RDF were launched in 2015. The result for RDF was too large to process it together with the other metadata formats. Hence, we produced the timeline for RDF separately. The timelines across all repositories reveal a steadily increasing number of datasets being published at \emph{GBIF}, \emph{Dryad}, \emph{Zenodo} and \emph{Figshare}. For \emph{PANGAEA}, the timeline points to a constant number of published datasets of around 10,000 datasets a year apart from an initial release phase between 2003 and 2007.

\begin{table}[t]
\small
	\begin{adjustwidth}{-1.0in}{0in}
	\caption{{ \bf Total number of datasets parsed per data repository and metadata schema. The numbers in brackets denote the number of datasets used for the analysis. All datasets were harvested and parsed in May 2019.}}
	\begin{center}
     \begin{tabular}{|p{3cm}|p{2.5cm}|p{2.5cm}|p{2.5cm}|p{3cm}|p{3cm}|}
	  \hline
		 \textbf{Metadata Schema}& \textbf{Dryad} & \textbf{PANGAEA} & \textbf{GBIF} & \textbf{Zenodo} & \textbf{Figshare} \\
		\hline \hline
		\emph{OAI-DC} & 186951 (142329)& 383899 (383899) & 44718 (42444) & 255000 (255000)& 3128798 (3128798)\\ \hline
		\emph{QDC} & & & & & 1718059 (1718059)\\ \hline		
		\emph{RDF} & 186955 (142989) & & & & 3157347 (3157347)\\ \hline
		\emph{DATACITE} & & & & 1268155 (1268155) &\\ \hline
		\emph{DATACITE3} & & 383906 (383906) & & 1268232 (1268232) &\\ \hline
		\emph{OAI-DATACITE} & & & & 1266522 (1266522) & 3134958 (3134958)\\ \hline		
		\emph{OAI-DATACITE3} & & & & 1268679 (1268679) &\\ \hline		
		\emph{DATACITE4} & & & & 1268262 (1268262) &\\ \hline		
		\emph{EML} & & & 44718 (42444) & &\\ \hline
		\emph{DIF} & & 383899 (383899) & & &\\ \hline
		\emph{ISO19139} & & 383899 (383899) & & &\\ \hline
		\emph{ISO19139.iodp} & & 383899 (383899)& & &\\ \hline
		\emph{PAN-MD} & & 383899 (383899) & & &\\ \hline
	\end{tabular}
	\label{DataRepoStatistics}
	\end{center}
	\end{adjustwidth}
\end{table}

\begin{figure}[t]
\begin{adjustwidth}{-.5in}{-.5in}  
\centering
\includegraphics[width=0.43\textwidth]{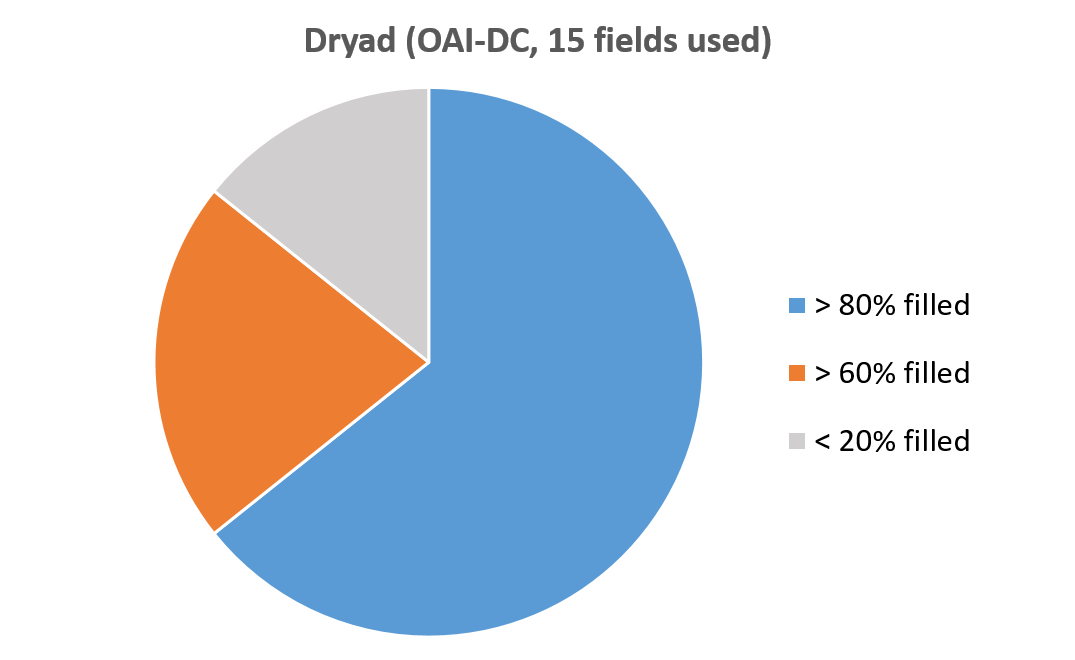}\hfill
\includegraphics[width=0.4\textwidth]{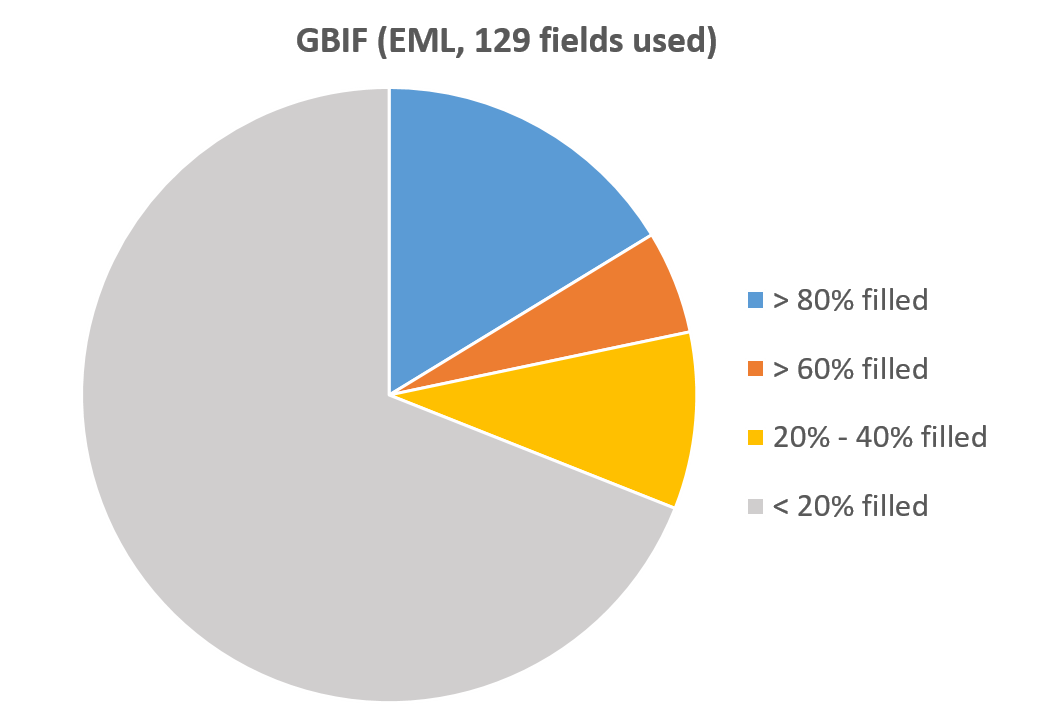}\hfill
\includegraphics[width=0.4\textwidth]{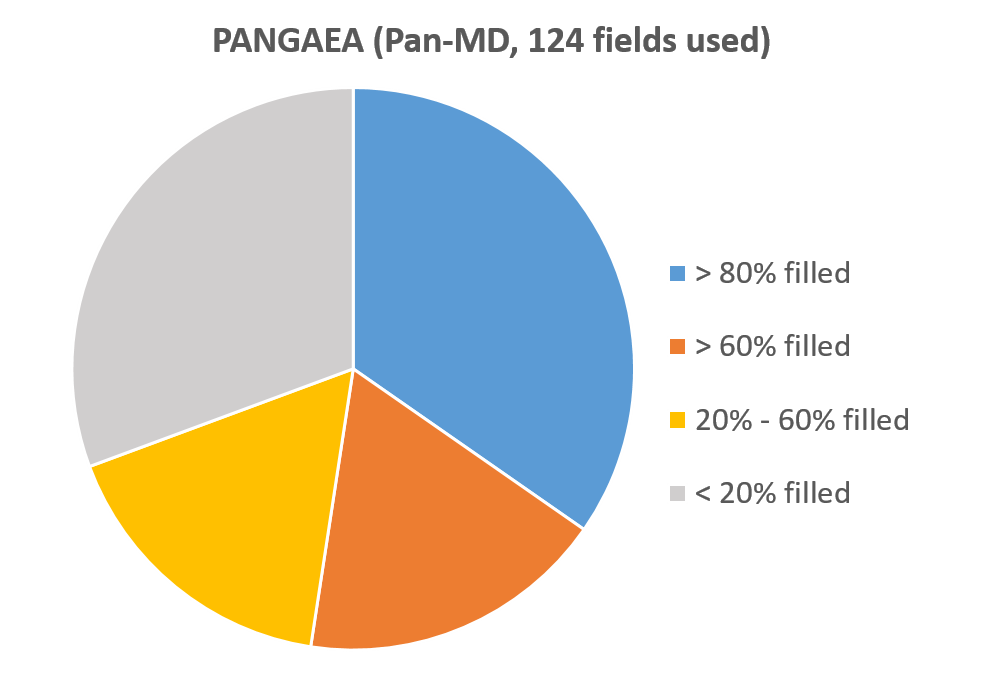}\hfill
\includegraphics[width=0.40\textwidth]{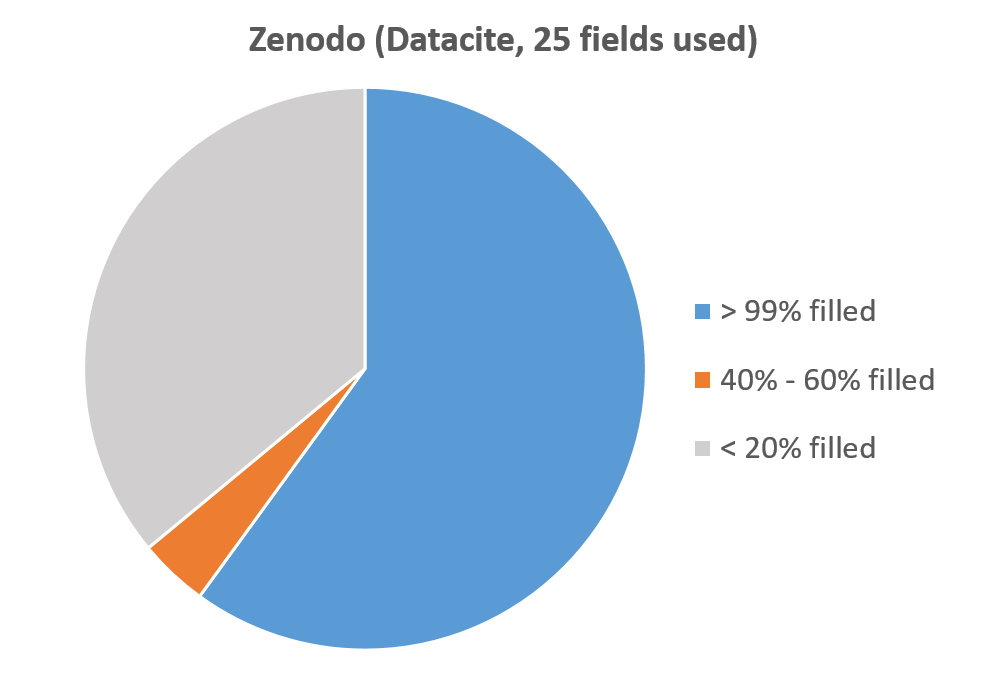}\hfill
\includegraphics[width=0.38\textwidth]{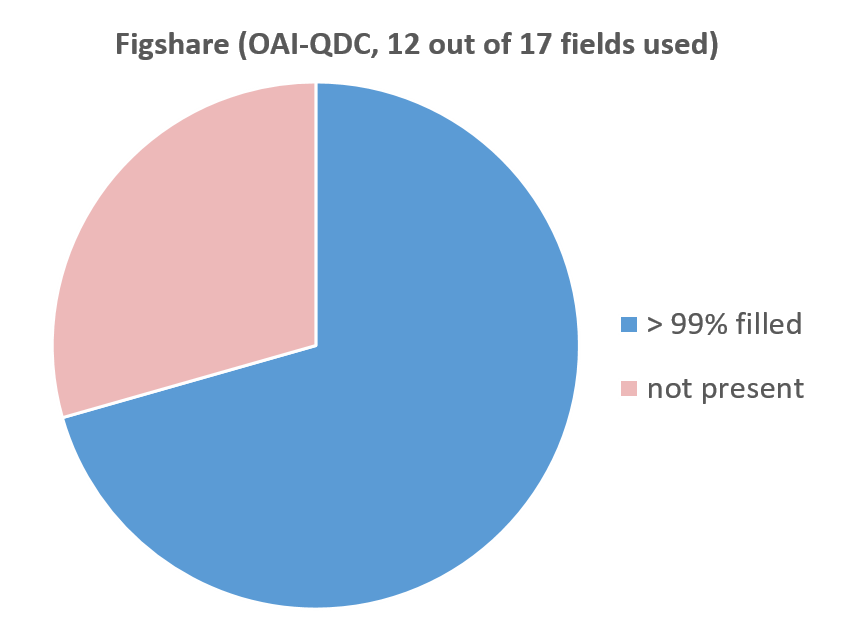}\hfill
\caption{Metadata field usage in all data repositories evaluated.} \label{fig:field_usage}
\end{adjustwidth}
\end{figure}

\paragraph{Metadata Field Usage}
Figure \ref{fig:field_usage} presents how many metadata elements of the best matching standard were filled. The individual results per data archive are available in our repository as supplementary material. \emph{Dryad} used 9 out of 15 available metadata fields from OAI-DC very often (> 80\%) including important fields such as \texttt{dc:title}, \texttt{dc:description} and \texttt{dc:subject}. \texttt{dc:publisher} and \texttt{dc:contributor} were provided in less that 20\%. For \emph{GBIF}, the EML standard does not provide a fixed number of core elements. Hence, we analyzed the 129 available fields. Most of them (89 elements) were not filled, e.g., fields describing taxonomic information. Data about author, title and description were provided in more than 80\%. The general field \texttt{eml:keyword} was used in around 20\%. Out of 124 used fields in \emph{PANGAEA's} Pan-MD format,  43 fields were filled in more than 80\% of the harvested metadata files including information on the author, project name, coordinates, data parameters and used devices. Fields that were less filled are supplementary fields, for instance for citation, e.g., volume, pages. For \emph{Zenodo}, all required fields in DataCite (identifier, creator, title, publisher, publication year) were always filled. In addition, title, rights and descriptions as well as resource type were also provided in more than 99\% of the analyzed metadata files. However, in only 45\% of the metadata files, keywords (\texttt{subject}) were present. \emph{Figshare} used only 12 out of 17 available fields of QDC, but these fields were always filled.

\paragraph{Category - Field - Match}
Per data repository and metadata format, we computed charts that visualize which field was filled at what percentage rate and if they correspond to the categories introduced in Section ``\nameref{sec:questions}''. Table \ref{tableDataRepoResults} presents a summary of all data repositories and their best matching standard. The individual results per repository and the concrete field-to-category mapping are available in our repository.

Temporal expressions (TIME) and information about author and/or creator (PERSON) were mostly provided in all repositories. Apart from $PANGAEA$, repositories mainly provided the publication date and only partially added information about when the data was collected. Information about data type and formats was also contained in all metadata files apart from \emph{GBIF}. The identified search categories were partially covered by two repositories. $GBIF$ with EML reflects most of the categories, but fields that correspond to ENVIRONMENT, ORGANISM, DATA TYPE and METHOD were rarely filled.
Metadata files in $PANGAEA's$ repository-developed standard Pan-MD always contained information on data parameters (QUALITY) and geographic locations. In most cases, research methods and devices used were also given. \emph{Dryad} provided geographic information (LOCATION) in its \texttt{dc:coverage} field in at least 60\%.

{\renewcommand{\arraystretch}{1.5}%
\begin{table}[ht!]
\scriptsize
\begin{adjustwidth}{-1.2in}{0in} 
\caption{
{\bf Comparison of data repositories and their best matching standard with the information categories. The categories are sorted by frequency. The asterisk denotes the categories with an agreement less than 0.4}}
\begin{tabular}{|p{2.5cm}|p{0.6cm}|p{0.8cm}|p{0.6cm}|p{0.6cm}|p{0.6cm}|p{0.8cm}|p{1.1cm}|p{1.6cm}|p{0.6cm}|p{0.6cm}|p{0.6cm}|p{2.1cm}|p{0.8cm}|}
\hline
& \rotatebox[origin=c]{90}{\textbf{Environment}}   & \rotatebox[origin=c]{90}{\textbf{Quality*}} & \rotatebox[origin=c]{90}{\textbf{Material}} & \rotatebox[origin=c]{90}{\textbf {Organism}} & \rotatebox[origin=c]{90}{\textbf{Process}} &  \rotatebox[origin=c]{90}{\textbf{Location}} &  \rotatebox[origin=c]{90}{\textbf{Data Type*}} &  \rotatebox[origin=c]{90}{\textbf{Method*}} & \rotatebox[origin=c]{90}{\textbf{Anatomy*}} &  \rotatebox[origin=c]{90}{\textbf{Human Intervention*}} &  \rotatebox[origin=c]{90}{\textbf{Event*}} &  \rotatebox[origin=c]{90}{\textbf{Time}} & \rotatebox[origin=c]{90}{\textbf{Person}}\\ \thickhline
\hline
$GBIF$ ($EML$)  & \cellcolor{Yellow} (3\%) & \cellcolor{YellowOrange} & \cellcolor{YellowOrange} & \cellcolor{Yellow} (11\%) & \cellcolor{YellowOrange} & \cellcolor{Yellow} (35\%) & \cellcolor{Yellow} (8\%) & \cellcolor{Yellow} (18\%) & \cellcolor{YellowOrange}& \cellcolor{YellowOrange} & \cellcolor{YellowOrange} & \cellcolor{Yellow} (publication Date - 100\%, collection Date - 10\%) & \cellcolor{Yellow} ($>$90\%)\\ \hline
$Dryad$ ($OAI-DC$)  & \cellcolor{YellowOrange} & \cellcolor{YellowOrange} & \cellcolor{YellowOrange} & \cellcolor{YellowOrange} & \cellcolor{YellowOrange} & \cellcolor{Yellow} (60\%) & \cellcolor{Yellow} & \cellcolor{YellowOrange} & \cellcolor{YellowOrange}& \cellcolor{YellowOrange} & \cellcolor{YellowOrange} & \cellcolor{Yellow} (publication Date) & \cellcolor{Yellow} (80\%)\\ \hline
$PANGAEA$ ($Pan-MD$)  & \cellcolor{YellowOrange}& \cellcolor{Yellow} ($>$90\%) & \cellcolor{YellowOrange} & \cellcolor{YellowOrange} & \cellcolor{YellowOrange} & \cellcolor{Yellow} (100\%) & \cellcolor{Yellow} (100\%) & \cellcolor{Yellow} (Devices used - 90\%, research methods - 65\%) & \cellcolor{YellowOrange}& \cellcolor{YellowOrange} & \cellcolor{YellowOrange} & \cellcolor{Yellow} (publication Date - 100\%, collection Date - 80\%) & \cellcolor{Yellow} (100\%)\\ \hline
$Zenodo$ ($OAI-Datacite$)  & \cellcolor{YellowOrange}& \cellcolor{YellowOrange} & \cellcolor{YellowOrange} & \cellcolor{YellowOrange} & \cellcolor{YellowOrange} & \cellcolor{YellowOrange} & \cellcolor{Yellow} (100\%) & \cellcolor{YellowOrange} & \cellcolor{YellowOrange}& \cellcolor{YellowOrange} & \cellcolor{YellowOrange} & \cellcolor{Yellow} (publication Date) & \cellcolor{Yellow}(100\%)\\ \hline
$Figshare$ ($QDC$)  & \cellcolor{YellowOrange}& \cellcolor{YellowOrange} & \cellcolor{YellowOrange} & \cellcolor{YellowOrange} & \cellcolor{YellowOrange} & \cellcolor{YellowOrange} & \cellcolor{Yellow} (100\%) & \cellcolor{YellowOrange} & \cellcolor{YellowOrange}& \cellcolor{YellowOrange} & \cellcolor{YellowOrange} & \cellcolor{Yellow} (publication Date) & \cellcolor{Yellow}(100\%)\\
\hline
\end{tabular}
\\
\begin{flushleft}
\textbf{Table key}
\\
\begin{tabular}{llll}
\textcolor{YellowOrange}{$\blacksquare$} & Unspecific (generic element) &
\textcolor{Yellow}{$\blacksquare$} & Available (one or more elements)\\
\multicolumn{4}{c}{Amount in brackets denotes the percentage the element is filled.}
\end{tabular}
\end{flushleft}
\label{tableDataRepoResults}
\end{adjustwidth}
\end{table}
}

\paragraph{Content Analysis:}
Table \ref{TableTopKeywords} presents the Top5 keywords in the metadata field \texttt{dc:subject} for all repositories sorted by their frequencies. The full keyword lists are available in our repository.

For \emph{GBIF} datasets, in 81\% an empty \texttt{dc:subject} field was returned. \emph{Zenodo's} metadata provided keywords  in 52\% of the inspected cases. None of the repositories seem to consider upper and lower cases. For several terms, different spellings resulted in separate entries. Numerous keywords in \emph{PANGAEA} and \emph{Dryad} reveal that both repositories host marine data. \emph{PANGAEA}'s list mainly contains data parameters measured and used devices. In contrast, \emph{Dryad's} list indicates that terrestrial data are also provided. For instance, the lower ranked terms contain entries such as Insects (1296) (insects (180)) or pollination (471) (Pollination (170)). Geographic information , e.g., California (9817), also occured in \emph{Dryad's} \texttt{dc:subject} field. \emph{Zenodo's} and \emph{Figshare's}  keyword lists contain numerous terms related to collection data. We checked the term `Biodiversity' in both repositories in their search interfaces on their websites. It turned out that the \emph{Meise Botanic Garden} ({\url{https://www.plantentuinmeise.be}) provided large collection data in \emph{Zenodo}. Hence, each occurrence record counted as a search hit and got the label `Biodiversity'. We also discovered that \emph{Figshare} harvests \emph{Zenodo} data which also resulted in high numbers for \emph{Figshare} and the keyword `Biodiversity' (219022).


\begin{table}[ht!]
\small
\begin{adjustwidth}{-1.2in}{0in} 
\caption{
{\bf Top5 keywords and their frequencies in the metadata field \texttt{dc:subject}.
}}
\begin{tabular}{|p{2cm}|p{3cm}|p{3cm}|p{3cm}|p{3cm}|p{3cm}|}
\hline
 & \textbf{Pangaea}  & \textbf{GBIF}  & \textbf{Dryad} & \textbf{Zenodo} & \textbf{Figshare} \\ \hline
& water (201102) & Occurrence (6510), occurrence (46) & Temperature (16652), temperature (15916) & Taxonomy (459877), taxonomy (105)& Medicine (1057684), medicine (240)\\ \hline
& DEPTH (198349), Depth(71916) & Specimen	(3046), specimen (22) & Integrated Ocean Observing System (16373) & Biodiversity (458336), biodiversity (8593)& Biochemistry (1015906), biochemistry (92) \\ \hline
& Spectral irradiance	(175373) & Observation	(2425), observation (24)& IOOS (16373) & Herbarium (270110), herbarium (91) & Biological Sciences not elsewhere classified (983829)\\ \hline
& DATE/TIME	(128917) & Checklist (589), checklist (43) & Oceanographic Sensor Data (15015) & Terrestrial (269900), terrestrial (177) & Chemical Sciences not elsewhere classified (842865)\\ \hline
& Temperature (118522), temperature (50) & Plantas (368), plantas (42) & continental shelf (15015) & Animalia (205242), animalia (261) & Biotechnology (792223), biotechnology (23978)\\ \hline \hline
empty \texttt{dc:subject} & 0 & 38296 & 15436 & 705730 & 0\\
\hline \hline
\end{tabular}
\label{TableTopKeywords}
\end{adjustwidth}
\end{table}


In a second analysis, we investigated which kinds of entity occur in descriptive metadata fields. As the processing of textual resources with NLP tools is time-consuming and resource-intensive, we selected a subset of datasets. We limited the amount to 10,000 datasets per repository. Table \ref{tableFilter} presents the filter strategies. For \emph{PANGAEA} and \emph{GBIF}, we randomly selected 10,000 datasets as they are domain-specific repositories for which all data are potentially relevant for biodiversity research. For \emph{Dryad}, the filter consists of a group of relevant keywords, and for \emph{Zenodo} and \emph{Figshare} we used the keyword `Biodiversity'. Due to the large amount of collection data with the keyword `Biodiversity', we are aware that this filter strategy might have led to a certain bias in the selected data.

\begin{table}[ht!]
\small
\begin{adjustwidth}{-1.2in}{0in} 
\caption{
{\bf Filter strategies used per data repository to select 10,000 datasets. The number in brackets denotes the total number of available datasets in OAI-DC format at the time of download (October/November 2019).}}
\begin{tabular}{|p{2cm}|p{2.5cm}|p{2.5cm}|p{5cm}|p{2.5cm}|p{2.5cm}|}
\hline
 & {\bf Pangaea}  & {\bf GBIF}  & {\bf Dryad} & {\bf Zenodo} & {\bf Figshare} \\ \hline
filter strategy &
10000 randomly selected (388254) &
10000 randomly selected GBIF (46954)&
10000 randomly with keywords: biodiversity, climate change, ecology,
insects, species richness, invasive
species, herbivory,
pollination, endangered species,
ecosystem functioning, birds (149672)&
10000 randomly with keyword: Biodiversity (1467958)&
10000 randomly with keyword: Biodiversity (3602808)
 \\ \hline
\end{tabular}
\label{tableFilter}
\end{adjustwidth}
\end{table}

Per data repository, we processed the selected 10,000 files with two open-source taggers of the text mining framework GATE \cite{GATE}. Named Entities such as Person, Organization and Location were obtained with the ANNIE pipeline \cite{ANNIE}, and Organisms were obtained from the OrganismTagger \cite{Naderi2011}. The results are presented in Table \ref{tableContentAnalysis}. Unfortunately, the OrganismTagger pipeline aborted for \emph{PANGAEA} and \emph{Zenodo}, but in around 12\% \emph{GBIF} files, 36\% \emph{Dryad} files and 85\% \emph{Figshare} files `Organism' annotations were created. Probably, the number of `Organism' annotations in \emph{Figshare} files is that high due to the mentioned bias towards collection data. The number of `Organism' annotations in \emph{GBIF} files is low since datasets mostly describe the overall study and do not contain concrete species names but rather broader taxonomic terms such as `Family' or `Order'. A large number of `Location' annotations were extracted for files from \emph{PANGAEA} (91\%) and \emph{Figshare} (~100\%). `Person' and `Organization' annotations are largely presented in \emph{PANGAEA} (~51\%) and \emph{GBIF} files (~74\%).

The text mining pipelines were originally developed and evaluated with text corpora and not sparse datasets. Hence, the results might contain wrong (false positive) annotations. However, the results indicate that NLP tools can support the identification of biological entities. That could be an approach for generalist repositories to additionally enrich metadata. All scripts and the final results are available in our repository.

\begin{table}[ht!]
\small
\begin{adjustwidth}{-0in}{0in} 
\caption{
{\bf NLP analysis: Number of datasets with Named Entities (out of 10,000 processed files in a reduced OAI-DC schema) per repository. Each file contains a subset of the original metadata, namely, \texttt{dc:title}, \texttt{dc:description}, \texttt{dc:subject} and \texttt{dc:date}.
}}
\begin{tabular}{|p{1.7cm}|p{1.7cm}|p{1.7cm}|p{1.7cm}|p{1.7cm}|p{1.7cm}|}
\hline
 & {\bf Pangaea}  & {\bf GBIF}  & {\bf Dryad} & {\bf Zenodo} & {\bf Figshare} \\ \hline
Organism & N/A (pipeline aborted) & 1183 & 3603 & N/A (pipeline aborted) & 8542 \\ \hline
Location & 9111 & 5642  & 3530 & 4641 & 9978\\ \hline
Person \& Organization & 5048 & 7355 & 657 & 192 & 1645\\ \hline
\end{tabular}
\label{tableContentAnalysis}
\end{adjustwidth}
\end{table}

\section*{D - Discussion}
\label{sec:discussion}

In this study, we explored what hampers dataset retrieval in biodiversity research. The following section summarizes our findings and outlines a proposal on how to bridge the gap between search interests in biodiversity and given metadata. We also highlight challenges that are not fully resolved yet.

\subsection*{Research Contributions}

\paragraph{Scholarly Search Interests in Biodiversity Research}
In order to understand what biodiversity scholars are interested in, we gathered 169  questions, identified biological entities and classified the entities in 13 information categories. In the subsequent evaluation with domain experts, five categories were verified and  can be considered as important information needs in biodiversity research. That includes information about habitats, ecosystems, vegetation (ENVIRONMENT), chemical compounds, sediments and rocks (MATERIAL), species (ORGANISM), biological and chemical processes (PROCESS). Further categories being mentioned very often are information about data parameters (QUALITY) and the nature or type of data resources (DATA TYPE). Usually, the latter is an outcome of a certain research method. However, the naming should be discussed in the research community as the comprehensibility of these categories were fair, only.

\paragraph{Comparison of Metadata Standards and User Interests}
We selected 13 metadata standards used in the Life Sciences from \emph{re3data}, and we analyzed whether the elements of the metadata schemes reflect the identified information categories.

Elements of general standards cover the categories to some extent, only.  LOCATION and DATA TYPE are the sole information that can be explicitly described with metadata fields of general standards such $DublinCore$ or $DataCite$. Further elements are focused on  information less relevant for search such as data creator, contributor (PERSON), collection or publication data (TIME), and license information. All this information is important for data reuse and data citation and needs to be part of the metadata. However, if the dataset is not findable, it can not be cited. As a general standard, $DataCite$ provides many more fields and attributes to describe personal data, time and geographic information. Therefore, it should be provided in addition to $DublinCore$.

There are numerous discipline-specific standards that describe search interests quite well. For instance, $EML$, $DarwinCore$ and $ABCD$ provide elements to describe environmental information, species, methods, and data parameters. \emph{ISA-Tab}, a framework for genome data and biological experiments covers all important search categories. 
The only drawback is that it takes time for scholars and/or data curators to fill in all these fields. In some standards such as $ABCD$ more than 1000 elements are available.
With our work, we aim to provide insights on what scholars are actually interested in when looking for scientific data. We believe that our results could serve as a good start for discussions in the respective research communities to define core elements of discipline-specific standards that are not only focused on data citation but also broaden the view on search interests.

\paragraph{Metadata Analysis of Selected Data Repositories}
We selected 5 repositories from \emph{Nature's} list of recommended data archives and analyzed the metadata provided in their OAI-PMH interfaces. We wanted to know what metadata standards are used in common data repositories in the Life Sciences and how many elements of the standard are actually filled.

We figured that generalist repositories such as $Dryad$, $Zenodo$ and $Figshare$ tend to use only general standards such as $DublinCore$ and $DataCite$. Even when using simple standards the repositories did not fully use all provided elements. Furthermore, the ones utilized are not always filled. That hampers successful data retrieval. Most repositories seem to be aware of that problem and enhance metadata with numerous keywords in generic fields such as \texttt{dc:subject}. Discipline-specific repositories, e.g., $GBIF$ and $PANGAEA$ are more likely to provide domain-related standards such as $EML$ or \emph{Pan-MD}. That supports an improved filtering in search, however, it does not guarantee that the fields are always filled. In \emph{GBIF's} case, we are aware that we could not provide a full picture as we did not analyze the occurrence records. Here, only a deeper analysis of the provided fields in the search index would deliver more answers. However, that would require technical staff support as the access to search indices is limited.


\subsection*{Suggestions to Bridge the Gap}

In this subsection, we outline approaches to overcome the current obstacles in dataset search applications based on our findings from the preceding sections. Table \ref{recommendations} presents checklists for data repositories and scholars that in the following are discussed in detail.

\begin{table}[!ht]
\begin{adjustwidth}{-1.2in}{0in} 
\centering
\caption{
{\bf Recommendations for data repositories and scholars to create rich metadata}}
\begin{tabular}{p{9.5cm}p{9.5cm}}
\rowcolor{bananamania}
\textbf{For Data Repositories} & \textbf{For Scholars} \\ [5pt]
\rowcolor{bananamania}
1.) Keep metadata diversity by offering domain-specific standards.&
1.) If available, select a domain-specific repository where possible (as recommended by \emph{Nature}). \\  [5pt]
\rowcolor{bananamania}
2.) Use metadata standards that include the information categories identified in Section A - Information Needs to cover potential search interests and make your data findable. &
2.) Check if appropriate, discipline-specific metadata standards are offered to describe your data.\\ [15pt]
\rowcolor{bananamania}
3.) Extend existing standards where necessary, preferably get in touch with metadata standard consortia. &
3.) Fill in all applicable metadata fields and use appropriate terms (if possible from controlled vocabularies).\\ [5pt]
\rowcolor{bananamania}
4.) Fill in all metadata fields. If possible use controlled vocabularies to describe the data, preferably Linked Open Data. &
4.) If your data is available in search, check if you can find it with various search terms.\\ [5pt]
\rowcolor{bananamania}
5.) Enrich your metadata with entities from \emph{schema.org} or \emph{bioschemas.org}.  &
5.) Contact the data repository if you notice issues in your data description or presentation. \\ [5pt]
\rowcolor{bananamania}
6.) In addition to explicit information, attempt to extract implicit information from metadata fields that contain longer textual resources, e.g., title, description and abstract. &
\\ [5pt]
\end{tabular}
\label{recommendations}
\end{adjustwidth}
\end{table}

\paragraph{For Data Repositories}
Adherence to the FAIR principles, long-term data preservation and the creation of citable and reusable data are main targets of all data repositories. Therefore, a strong focus of data archives is on generating unique identifiers, linking the metadata to their primary data and publications.
Less considered is the perspective of dataset seekers.
Hence, we propose the following improvements to enhance dataset retrieval.\\

\noindent
\emph{Keep metadata diversity}: Scientific data are very heterogeneous. This diversity can not be reflected in one generic metadata standard. Thus, it is highly recommended to use different domain-specific standards considering the requirements from various research disciplines.\\

\noindent
\emph{Use proper metadata fields}: If search interests are explicitly mentioned in metadata, conventional search techniques are able to retrieve relevant datasets. Providing possible search terms in generic keyword fields supports dataset retrieval in a full-text search but does not allow proper category-based facet creation. 
Therefore, using proper metadata fields covering potential search interests greatly enhances dataset retrieval and filtering.

In addition, metadata need to have a unique identifier and should answer the W-questions including information on how the data can be re-used. That comprises information on data owner, contact details and citation information.\\

\noindent
\emph{Extend standards}: Metadata standards are developed and adopted from large organizations or research communities for a specific purpose or research fields. They also discuss extensions of new fields or changes of existing elements. If the given fields are not sufficient for particular requirements, the preferred way is to get in touch with the standard organization and to propose new fields or attributes. However, since these processes usually take a long time, it is sometimes unavoidable to extend a schema or to develop a new schema. In these cases, it would be a good scientific practice to give feedback to the standard organization why and how a schema has been changed or extended. That might influence the further development of standards and would counteract the creation of numerous repository-developed schemes.\\

\noindent
\emph{Use controlled vocabularies}: The questions that still remain and that have not been considered so far are how metadata fields are filled - by the data submitter, the data repository or by the system - and whether a controlled vocabulary is used for the keywords and the other metadata elements. When describing scientific data it is highly recommended to use controlled vocabularies or terminologies, in particular for important fields in search. If possible, Linked Open Data \cite{LinkedDataBook} vocabularies should be utilized to better link datasets, publications, authors and other resources. That supports data transparency, data findability and finally data reuse. In the Life Sciences, there are a variety of terminology providers. We provide a list of ontology and semantic service providers in our repository.\\


\noindent
\emph{Utilize schema.org}: Driven by \emph{Google} and the Research Data Alliance (RDA) Data Discovery Group, the enrichment of HTML with \emph{schema.org} (\url{https://schema.org}) entities became very popular in recent years. The enrichment helps to identify unique identifiers, persons, locations or time information in the HTML file. That supports external search engines or data providers to crawl the landing pages of search applications provided by the data repositories per dataset.
As the current schema.org entities do not fully reflect scientific search interests, more attention should be paid to initiatives such as \emph{bioschemas.org} (\url{https://bioschemas.org/}) that aims to expand schema.org on biological entities such as species and genes. That confirms and complements our recommendations for explicit metadata fields tailored to search interests. At the time of writing this paper, \emph{bioschemas.org} is still in draft mode. However, in the future, efforts like this will improve dataset retrieval significantly.\\

\noindent
\emph{Extract implicit information}: Apart from short information such as contact details, data type or location, metadata usually contain longer textual resources such as title, description and abstract. Most of them contain useful information for search and mention species observed or describe environments where data has been gathered. These resources could be used to extract implicit information and to automatically identify further relevant data.

\paragraph{For Scholars}
Documenting scientific data is a disliked task that also takes time. Therefore, scholars attempt to minimize the effort on describing their data and are pleased when data repositories offer not too many fields to fill in for data submission. However, scholars are responsible to properly document their data so that other researchers are able to find and reuse it. Hence, each scholar should carefully and thoroughly describe the produced research data. Based on our findings, we summarize what should be considered when submitting scientific data to a data repository.\\

\noindent
\emph{Prefer domain-specific repositories:} As generalist repositories tend to offer only general metadata standards for data description, preference should be given to domain-specific data archives. This is also recommend by highly influential journals such as \emph{Nature} \cite{NatureRepoList}. Another advantage is that repositories that are familiar with the research domain might give more qualitative feedback on the submitted data descriptions.\\

\noindent
\emph{Use domain-specific metadata standards:} Even when selecting a domain-specific data repository, it does not guarantee that archives use proper metadata standards. Scholars are advised to know at least a few appropriate standards for their research field and to ask the repository if one of these standards are supported if not stated anywhere.\\

\noindent
\emph{Fill in all relevant fields with controlled vocabularies:} All relevant metadata fields should be filled in. That enhances the chance that datasets are retrieved. When describing the data, scholars should attempt to use controlled vocabularies. As this is a new procedure in data submission, it is currently not supported by all data repositories. However, if it is available, it is recommended to use the terminologies given and not to describe the data with one's own words.\\

\noindent
\emph{Search for your data:} Once the data is available in the repositories' search application, scholars are advised to check if they can find their data with various search terms. They should also review whether the data are accessible and all displayed information are correct. It is also recommended to repeat this checking from time to time as repositories might update or extend data presentations and/or metadata schemes used.\\

\noindent
\emph{Get in touch with the repository:} If scholars notice anything concerning their data, they should contact the archive. The staff at the repositories are probably grateful if attentive scholars give feedback on their submitted data or detect issues that hampers dataset retrieval.

\subsection*{Challenges}
As stated in our summary of the question analysis, the outcomes in Section ``\nameref{sec:questions}'' are not a complete picture of search interests but only serve as a start for discussions with biodiversity researchers to further identify possible search categories.

Controlled vocabularies can only be used if appropriate terminologies exist. This is not the case for all topics. While there are numerous vocabularies for species, to the best of our knowledge, there is no vocabulary that allow the description of research methods and results. Scientific data types are also less considered in existing terminologies.

Another challenge lies in the automatic identification of relevant search topics in metadata. The text mining community has already developed various taggers and pipelines to extract organisms \cite{OrganismTagger2011}, chemistry items \cite{GATE2011} or genes \cite{McDonald2005} from text. These annotations can support automatic facet or category creation. However, for important categories such as habitats, data parameters, biological and chemical processes or research methods, taggers are still missing. In order to increase semantic linkage of datasets and other resources such as publications, authors and locations, it would be a great benefit if the annotations also contain URIs to resources in controlled vocabularies. Then, dataset retrieval could be expanded on semantically related terms such as synonyms or more specific or broader terms.

An important point, however, are not standards, systems or vocabularies, but scholars themselves. Scholars need to be aware that thorough data descriptions are part of a good scientific practice. In order to preserve all kind of scientific data, independently of whether it has been used in publications or not, proper metadata in appropriate schemes are the key to successful dataset retrieval and thus, to data citation and data reuse. Data Repositories could offer data curation services to support scholars in describing research data and to encourage them to describe their data thoroughly. We are aware that it would require high efforts to introduce more domain-specific metadata schemes at generalist repositories; however, it would enhance dataset retrieval.

Computer science research can contribute to improvements for dataset search by developing methods and software tools that facilitate standard-compliant metadata provision ideally at the time of data collection, thus ensuring metadata standards to be actually used by data providers.


\section*{Conclusion}
\label{sec:conclusion}
Scholarly search interests are as diverse as data are and can range from specific information needs such as searches for soil samples collected in a certain environment to broader research questions inspecting relationships among species. Our findings reveal that these search interests are not entirely reflected in existing metadata. One problem are general standards that are simple and mainly contain information that support data citation. Actual search interests can only be represented if keywords and suitable search terms are provided in general, non-specific fields that are provided in most standards, e.g., \texttt{dc:subject} in \emph{DublinCore}. Most data repositories utilize these fields to enrich metadata with suitable search terms. However, if search interests are not explicitly given, facet creation, e.g., filtering over species or habitats, is more difficult. Full-text searches only return data if query terms match given keywords. On the other hand, even when scholars submit their data to a domain-specific repository that uses discipline-specific metadata standards, it does not guarantee that all search-relevant fields will be filled. 

Data findability, one of the four FAIR principles \cite{FAIRPrinciples}, at least partially relies on rich metadata descriptions reflecting scholarly information needs. If the information scholars are interested in is not available in metadata, the primary data can not be retrieved, reused and cited. In order to close this gap, we propose checklists for data archives and scholars to overcome the current obstacles. We also highlight remaining challenges. In our future work, we would like to focus on a machine-supported extraction of relevant search categories in metadata as well as an automatic filling of metadata fields from primary data. That will minimize the metadata creation process and will support scholars and data repositories in producing proper and rich metadata with semantic enrichment.

\section*{Acknowledgments}
We acknowledge the Collaborative Research Centre AquaDiva (CRC 1076 AquaDiva) of the Friedrich Schiller University Jena and the GFBio project (KO2209/13-2), both funded by the Deutsche Forschungsgemeinschaft (DFG). The authors would also like to thank the annotators and reviewers for their time and valuable comments.


\bibliographystyle{apalike}
\bibliography{dataSearch}

\end{document}